\documentclass[12pt,preprint]{aastex}

\newcommand{\ud}{\mathrm{d}}
\slugcomment{Version 7.0, May 16, 2005}

\shorttitle{Cometary UCHII Regions}
\shortauthors{Zhu et al.}

\begin{document}


\title{Mass Flows in Cometary UCHII Regions}


\author{Qing-Feng Zhu\altaffilmark{1}, John H. Lacy\altaffilmark{1}, Daniel T. Jaffe\altaffilmark{1}}
\affil{Astronomy Department, University of Texas at Austin,
    Austin, TX 78712}

\email{zhuqf@astro.as.utexas.edu}

\and

\author{Thomas K. Greathouse\altaffilmark{1}}
\affil{Lunar and Planetary Institute, 3600 Bay Area Boulevard, Houston, TX 77058-1113}

\and 

\author{Matthew J. Richter\altaffilmark{1}}
\affil{Department of Physics, University of California, Davis, CA 95616-8677}


\altaffiltext{1}{Visiting Astronomer at the Infrared Telescope Facility,
which is operated by the University of Hawaii under Cooperative Agreement no.
NCC 5-538 with the National Aeronautics and Space Administration, Office of
Space Science, Planetary Astronomy Program.}


\begin{abstract}
High spectral and spatial resolution, mid-infrared fine structure line
observations toward two ultracompact HII (UCHII) regions
(G29.96~-0.02 and Mon~R2) allow us to study the
structure and kinematics of cometary UCHII regions. 
In our earlier study of Mon~R2,
we showed that highly organized mass motions accounted for most of 
the velocity structure in that UCHII region.
In this work, we show that the
kinematics in both Mon~R2 and G29.96 are consistent with motion along an
approximately paraboloidal shell.
We model the velocity structure seen
in our mapping data and test the stellar wind bow shock model for 
such paraboloidal like flows. 
The observations and the simulation
indicate that the ram pressures of the stellar wind and
ambient interstellar medium cause the accumulated mass in the bow shock 
to flow along the surface of the shock.
A relaxation code reproduces the mass flow's velocity structure
as derived by the analytical solution. It further predicts that the pressure
gradient along the flow can accelerate 
ionized gas to a speed higher than that of the moving star. 
In the original bow shock model, the star speed relative to the ambient medium
was considered to be
the exit speed of ionized gas in the shell. 
\end{abstract}


\keywords{interstellar medium - ultracompact H~II regions - G29.96~-0.02 - Mon~R2~IRS1}


\section{Introduction}

Ultracompact HII (UCHII) regions form when massive OB stars ignite inside
dense molecular clouds and high energy UV photons ionize the surrounding
neutral material.
They have small sizes (${\lesssim}0.1$~pc), high electron densities
(${\gtrsim}10^4$~cm$^{-3}$), and high emission measures
($EM={\int}n_{i}n_{e}{\ud}{l}~{\gtrsim}~10^{7}$pc{\,}cm$^{-6}$)
\citep{wooC89b}.
The ionization increases the temperature (${\sim}10^{4}$K) and 
number density
($H_2 \to 2p^{+} + 2e^{-}$) causing 
the gas pressure inside the regions to be at least two orders
of magnitude higher than in the surrounding molecular gas.
UCHII regions should then expand at approximately the speed of sound in the
ionized gas ($\sim 10$~km{\,}s$^{-1}$) until they reach
pressure equilibrium with the surrounding material or break out
of their parent molecular clouds.
Because of this expansion, the
density and emission measure both should drop with time.
Once the emission measure drops below $10^7$~pc{\,}cm$^{-6}$,
these HII regions will no longer be considered ultracompact. 

A longstanding puzzle regarding UCHII regions involves their numbers and
their lifetimes.
If HII regions expand at 10 km~s$^{-1}$, they should remain ultracompact for
only $\sim 10^4$~yr.
Since this is $<$1\% of the lifetime of an OB star, the number of UCHII regions
should be $<$1\% the number of OB stars.
Radio interferometry observations confirm that the ratio of IRAS 
far-infrared (FIR) flux densities
provides an efficient way to select embedded OB
stars from the IRAS data \citep{wooC89b,chu90,garRMC93,kurCW94,mirRS94}.
However, counts of IRAS FIR sources with the spectral
characteristics of UCHII regions and the optically visible O stars in the solar
neighborhood show
that OB stars spend $\sim 10 - 20\%$ of their main-sequence lifetime 
embedded in molecular clouds \citep{wooC89a}.
Apparently, the lifetimes of UCHII regions are an order of 
magnitude larger than predicted by
the classical pressure-driven spherical expansion model.

The variety of observed UCHII morphologies, including spherical, shell-like,
cometary, and irregular, also needs to
be explained.
An understanding of the kinematics of the ionized
gas will help provide an explanation of both the morphologies
and the lifetime of UCHII regions.
It has been suggested that long lifetimes could result from some kind of
containment or gas replenishment mechanism
\citep{holJLS94,dysWR95,redWD96,wilDR96,redWD98}.
Various mechanisms would have different effects on
the morphologies and the kinematics of UCHII region, suggesting that
observations could distinguish between these mechanisms.

The well-organized overall appearance of cometary UCHII regions
make them good objects in which to study UCHII region 
kinematics and morphology.
One way to explain these objects is with a bow shock model, which suggests
the cometary structure of some UCHII regions is the result of the 
supersonic motion
of OB stars with high speed winds through molecular clouds
\citep{hugV76,wooC89b,vanMWC90,macVW91,vanM92}. 
Swept-up ambient gas and stellar wind material accumulate 
where the ram pressures associated with these two mass 
flows balance, resulting in a shell-like structure and 
a surface flow.
This model may not explain the lifetime of all
UCHII regions, but could provide a part of the solution to this problem. 

Because ultracompact HII regions are formed inside dense molecular clouds,
the extinction toward these objects usually
is very high, with a typical line of sight extinction of 
$A_v=30-50$ \citep{hanLR02}. Thus, optical techniques
used in diffuse HII region observations are inappropriate
for studies of UCHII regions. 
Low extinction at infrared and radio wavelengths makes these
wavelength bands more suitable for studies of UCHII regions.
Hydrogen radio recombination line observations toward UCHII regions
have been carried out extensively \citep{garRM85,kimK01,araHCK02}.
These observations reveal that thermal motions cannot account for the
linewidths of many UCHII regions. However, significant thermal broadening
makes it hard to study the velocity structure in these regions  
using hydrogen recombination lines because of hydrogen's low atomic mass 
\citep[and references therein]{jafM99,sewCKGH04,depWMDGK04}. 
The thermal line width of hydrogen recombination lines is
${\Delta}{\upsilon}_{FWHM}{\simeq}21.4~$km{\,}s$^{-1}$ 
for gas with a temperature of $10^{4}$K.
This large thermal linewidth means that heavier 
ions are better probes of bulk motion;
the thermal broadening is 
only about $4.8~$km{\,}s$^{-1}$ for $Ne^{+}$ ions at the same temperature.
Mid-infrared ionic fine-structure lines have been used to probe
the structure and excitation of HII regions
\citep{becLTAGB81,lacBG82,takMWM00}, but
generally with too low spectral resolution to study gas motions.
To study kinematics and distinguish organized motions from turbulence and
thermal broadening of heavy ions,
one needs $\simeq 5~$km{\,}s$^{-1}$ velocity resolution.

We have begun a program of high spectral resolution observations of
mid-infrared fine structure line emission from
UCHII regions. By mapping a small sample of UCHII
regions of different morphologies at high
spatial resolution in [Ne~II]${\lambda}12.8{~\mu}$m,
[Ar~III]${\lambda}9.0{~\mu}$m,
[S~III]${\lambda}18.7{~\mu}$m and [S~IV]${\lambda}10.5{~\mu}$m, 
we can study 
the kinematics 
and physical conditions 
in UCHII regions
in order to better understand 
massive star formation and the relationship between these stars and their
surrounding environments.
Among these lines, 
the [Ne~II] line is 
particularly bright and is present 
in HII regions with a broad range of excitation, so it is well
suited for kinematic analysis.
We presented high spectral resolution observations of
one cometary UCHII region, Mon~R2~IRS1 (Mon~R2 hereafter), in \cite{jafZLR03}.
In the current paper, we present [Ne~II] observations of another 
cometary UCHII region, G29.96~-0.02 (G29.96 hereafter).
We also examine bow shock models in some detail 
and compare them to the observations of both G29.96 and Mon~R2.

In section {\bf 2}, we describe our mapping and
data reduction methods.
We present our [Ne~II] line emission observations toward G29.96 in
section {\bf 3}.
In section {\bf 4}, we
describe the kinematics of the bow shock model and our relaxation
method for simulating the formation of the surface flow.
We also introduce an additional acceleration, due to the pressure gradient
in the ionized gas along the flow.
In section {bf 5}, we compare the model predictions with the [Ne~II]
observations of G29.96 and Mon~R2. 
Finally, we discuss the existing problems of models for 
cometary UCHII regions in section~{\bf 6}.

\section{Observations}
\subsection{Instrument}

The [Ne~II] observations of G29.96 were carried out with TEXES
\citep[the Texas
Echelon Cross Echelle Spectrograph,][]{lacRGJZ02} on the NASA 3 meter 
Infrared Telescope Facility (IRTF) on Mauna Kea, in June, 2001.
TEXES is a high resolution (R $\leq$ 100,000) spectrograph operating at
mid-infrared (5-25 $\mu$m) wavelengths.
The slit, with a $1.4''$ width and a $11.5''$ length, was 
oriented north-south on the sky.
With this slit, we achieved a spectral 
resolution of ${\sim}4~$km{\,}s$^{-1}$ or R = 75,000 at $12.8{~\mu}$m.
Each pixel along the slit was about $0.36''$ on the sky,
which is a little smaller than half of the diffraction
limit ($0.88''$) of the IRTF at $12.8{~\mu}$m.
To map the regions of interest we stepped the telescope west 
to east across the objects without chopping.
Multiple, partially overlapping scans were needed 
to cover the entire region. For G29.96, each scan had a length of
$20''$ with a step size of $0.4''$. For Mon~R2, each scan was $45''$ long
with $0.7''$ steps \citep{jafZLR03}. Nine overlapping scans were made on
G29.96, and 20 scans were made on Mon~R2. 
Because the scans overlapped, the integration time at each point
in the map is approximately 18 seconds for G29.96 and 8 seconds for Mon~R2.

\subsection{Data Reduction}

The spectral-spatial datacubes produced from the scans 
were first reduced with a 
custom Fortran reduction program that performs general purpose corrections,
such as the correction of optical distortions, flat fielding, 
bad pixel masking
and cosmic ray spike removal \citep{lacRGJZ02}. 
Radiometric calibration was obtained from measurements of an ambient
temperature blackbody before each set of scans,
giving intensities in units of 
erg{\,}cm$^{-2}${\,}s$^{-1}${\,}sr$^{-1}${\,}(cm$^{-1})^{-1}$.
The uncertainties in the intensity are mostly systematic, and are
probably $\sim\pm20\%$.
The wavelength calibration was obtained from sky emission lines,
and is accurate to ${\sim}1~$km{\,}s$^{-1}$.
IDL scripts were used to do the sky background subtraction, 
multiple scan cross-correlation and combining, and datacube manipulation.
Positions safely off the object at both ends
of each scan allow us to interpolate the sky
emission at each step position. 
In doing the sky background subtraction, we assumed that the sky background 
varies linearly 
with time during the course of
a scan, based on the brightness of pixels off the object.
After sky emission subtraction, multiple scans of the same region
were cross-correlated, 
shifted, and added to make a complete datacube for the object.

\section{Results}

\begin{figure}
\epsscale{0.8}
\plotone{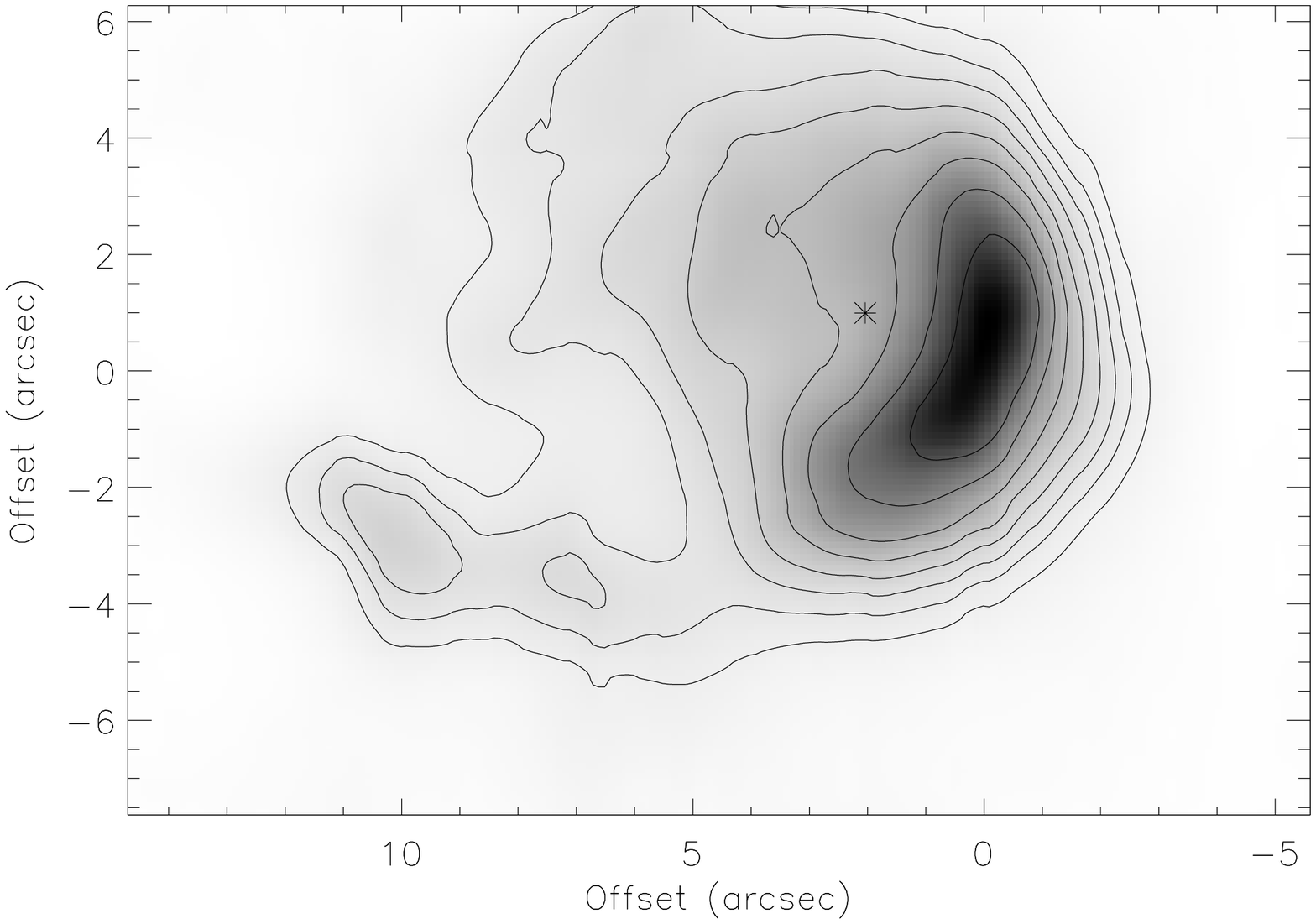}
\caption{Integrated line flux map of [Ne~II] line toward G29.96. The (0,0) postion marks
         the flux peak of the observation, corresponding to 
         $18^{h}46^{m}03^{s}.92, -02^{\circ}39'21''.9~ (J2000)$. The asterisk marks the position
         of the ionizing star from \cite{marBKTH03}. Contours are drawn at 
         $70\%$, $50\%$, $35\%$, $25\%$, $17.5\%$,
         $12.5\%$, $9\%$ and $6\%$ of the
         peak value ($1977~$ergs{\,}cm$^{-2}${\,}s$^{-1}${\,}sr$^{-1}${\,}(cm$^{-1})^{-1})$. 
\label{intmap}}
\end{figure}

\begin{figure}
\epsscale{0.8}
\plotone{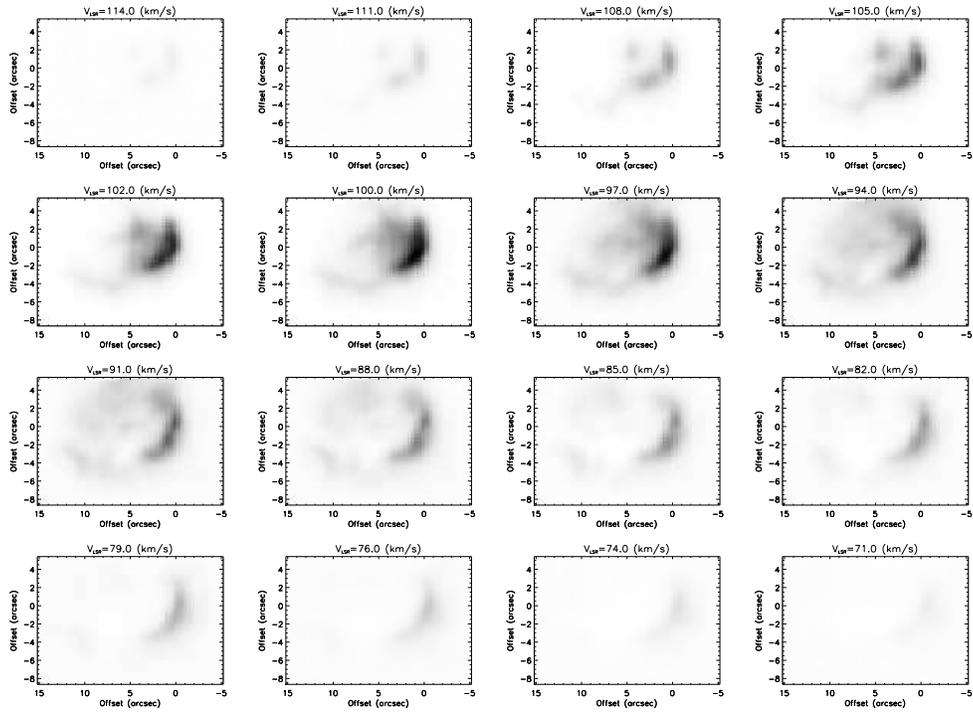}
\caption{Channel maps of G29.96. Corresponding V$_{LSR}$ is indicated in the plots.
\label{chanmap}}
\end{figure}

We show the integrated [Ne~II] line emission map for G29.96 in
Fig.~\ref{intmap}.
The overall cometary shape is apparent and is symmetric
about an axis at a position angle $\sim70^{\circ}$ east of north.
The brightest emission forms an arc perpendicular to the symmetry axis.
To the west of the arc, the line emission
drops rapidly from the peak value to the background level 
within $4''$.
On the east side of the bright arc, 
fainter emission extends over $13''$ to the east-northeast.
The change of the emission level is much more gradual and less uniform.
The east edge of the region is much fuzzier and broken
by a faint ``lane'' into two parts. 
The southern portion extends the curve of the emission arc out
$\sim6''$ and ends at a clump of ionized gas of size
$\sim3''\times3''$.
From our velocity channel maps (Fig.~\ref{chanmap}), we can see that  
it is the continuous extension of
the arc. 

Velocity channel maps of G29.96 (Fig.~\ref{chanmap}) show additional details.
A cavity is present at medium and lower velocity channels, 
from $\sim82~$km~s$^{-1}$ to $\sim94~$km~s$^{-1}$ .
The fainter structures on the east side of the region 
and the brighter arc form an almost complete ring around the cavity, 
although the symmetry axis of the elliptical ring is at a somewhat larger
position angle ($\sim120^{\circ}$) than that of 
the bright arc ($\sim70^{\circ}$).
A similar ring is seen in channel maps of Mon~R2 \citep{jafZLR03}, 
except that the 
major axis of the ring in Mon~R2 is nearly perpendicular to the symmetry axis
of its bright arc.
A compact emission peak 
centered at $V_{LSR}\sim108$~km{\,}s$^{-1}$
can be seen $\sim5''$ northeast 
of the peak of the arc in G29.96. 
The [Ne~II] line is generally broad, reaching
$\sim40~$km{\,}s$^{-1}$ near the bright arc.
The emission is typically bright near the velocity of the
ambient cloud \citep[$V_{LSR}\sim98~$km{\,}s$^{-1}$, 
obtained from single dish observations of rotational 
transitions of CS molecules,] 
[and references therein]{marBKTH03}, but the line is
significantly blueshifted ahead of the emission peak. The line center
reaches a $V_{LSR}\sim82~$km~s$^{-1}$ at the west edge of the region

\begin{figure}
\epsscale{0.5}
\plotone{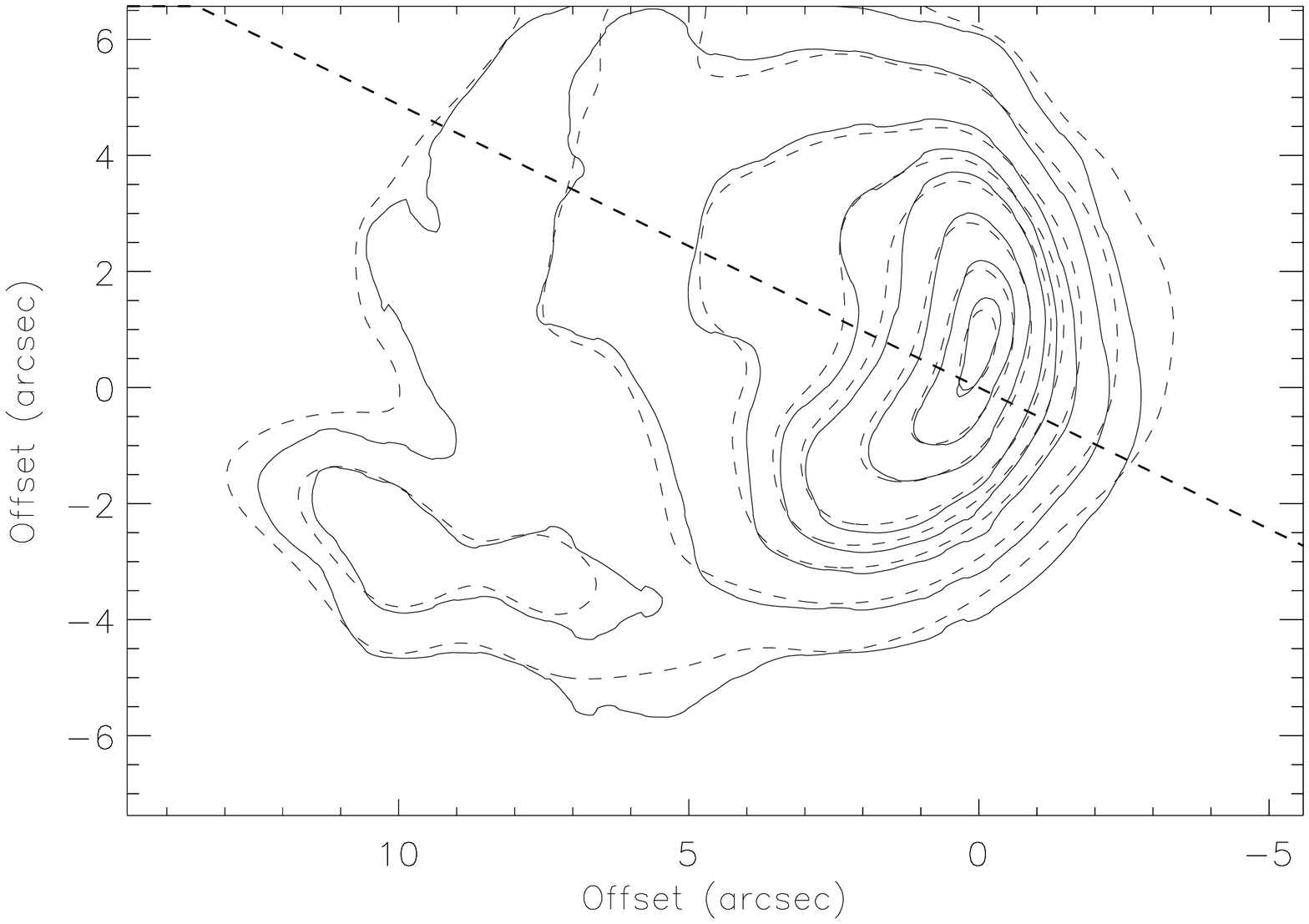}
\caption{Contours of [Ne~II] line map of G29.96 (solid line, 
         peak value $=1977~$erg{\,}cm$^{-2}${\,}s$^{-1}${\,}sr$^{-1}${\,}(cm$^{-1})^{-1}$) 
         overplotted on VLA 2~cm continuum observation \citep[dotted line, peak value $=0.078~$Jy/beam 
         with a $0.56''{\times}0.49''$ beam and a 100MHz bandwidth][]{feyGCV95}. 
         The VLA map has been smoothed with the ([Ne~II]) beam profile of 1.6'' FWHM (peak value $=0.037~$Jy/beam
         after the smoothing).
         Contours are plotted at 
         $5\%$, $10\%$, $20\%$, $30\%$, $40\%$, $60\%$, $80\%$, $95\%$ of the corresponding peak values.
         Coordinates are the offsets from the emission peak.
         The straight dashed line (p.a.=$64^\circ$)indicates where flux
         is taken for the flux distribution plot in Fig.~\ref{centflux}.  
\label{fluxcut}}
\end{figure}

\begin{figure}
\epsscale{0.5}
\plotone{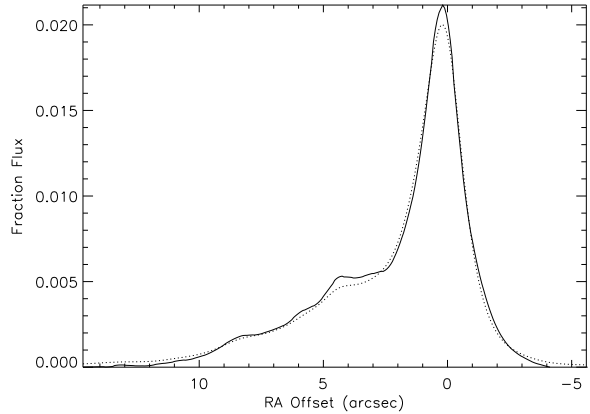}
\caption{Fluxes along the symmetry axis (shown in Fig.~\ref{fluxcut} with the dashed line) of G29.96~-0.02. 
The dotted line indicates the smoothed radio 2~cm flux, while the solid line indicates flux of [Ne~II]. The curves
are normalized to have equal integrals.
\label{centflux}}
\end{figure}

In Fig.~\ref{fluxcut}, we plot [Ne~II] line contours 
on the top of the VLA 2 cm continuum image of G29.96 \citep{feyGCV95}.
We first aligned the two maps, 
assuming they are emitted by the same ionized gas.
We then convolved the radio map with our beam.
The close resemblance of the maps after convolution suggests
that the much sharper edges seen in the
VLA image are the result of its higher spatial
resolution ($<0.56''$). 
Comparing a cross-cut along the symmetry axis (Fig.~\ref{centflux})
confirms that [Ne~II] and radio free-free flux are both
proportional to the emission measure of the region.
Radio recombination line observations at $0.62''$ and 4~km~s$^{-1}$ resolution
showed similar arc-like structure \citep{wooC91}. They also showed that 
the H76$\alpha$ line is broader 
in front of the radio continuum emission arc, where the velocity gradient
is also the highest.

\cite{marBKTH03} obtained a narrow-band
H$_2$~1-0~S(1) filter map and K band ($2.07-2.19{\mu}$m) 
spectra of the region with a resolution of R=8,000
along a $120''$ long slit along the symmetry axis.
Their map shows many local near-infrared emission
clumps together with the proposed ionizing star of 
the region lying $\sim2''.3$ northeast ($p.a.=64^{\circ}$) 
of the emission peak. 
We mark the star position with an asterisk in our map according 
to this offset (Fig. \ref{intmap}).
Their $Br\gamma$ linewidth ranges between
${\sim}42~$km{\,}s$^{-1}$ and ${\sim}62~$km{\,}s$^{-1}$ and shows 
significant blueshift in front of the emission arc, which is consistent with
our [Ne~II] observations. The fact that $Br\gamma$ line is systematically
broader than [Ne~II] along the slit supports the idea that [Ne~II] is a better
tracer of the ionized gas motions. 

\section{Bow Shock and Relaxation - A Physical Model for Parabolic Flow}

In this section, we present an extension of the analytic solution for
bow shocks using a numerical relaxation technique. We then use the resulting
gas flow as a template with which we can compare the 
observed kinematics of G29.96 and Mon~R2.

\subsection{Analytical Solution}

Bow shock models have been explored to explain the formation of 
cometary UCHII regions by several authors
\citep{vanMWC90,macVW91,vanM92,wil96}.
They describe a situation where a star with a strong stellar wind
and high UV luminosity moves inside a dense molecular cloud.
In the frame of reference of the star, the stellar wind and ambient
material collide and create a stationary shock region approximately
paraboloidal in shape in front of the star.
The standoff distance $r_{0}$ is determined by pressure balance
at the apex of the shell: 
\begin{eqnarray}
\rho_{w}v_{w}^2+n_{w}kT_{w}=\rho_{a}v_{*}^2+n_{a}kT_{a},
\end{eqnarray}
where $\rho_{w}=\frac{\dot{M}_{w}}{4{\pi}r_{0}^{2}v_{w}}$
is the stellar wind density at
the distance $r_{0}$ and $\rho_{a}=n_{a}{\mu}$ is the density of the
ambient medium. The corresponding number densities are $n_{w}$ and $n_{a}$.
Where the stellar wind and ambient medium collide near head-on, 
the ram pressures are much 
greater than the gas thermal pressure, thus the standoff
distance can be expressed as: 
\begin{eqnarray}
r_{0}=\sqrt{\frac{\dot{M}_{w}v_{w}}{4\pi\rho_{a}v_{*}^2}}
\end{eqnarray}
by neglecting gas thermal pressure \citep{wil96}.
The gas pressure becomes non-negligible 
far downstream from the star where the shell surface is roughly 
parallel to the stellar motion, resulting in reduced 
ram pressure from the
ambient medium.

For physical and mathematical simplicity, most models assume that
the stellar wind and ambient medium material are
well-coupled inside the shock region, and
radiative cooling is very efficient on the time scale of the ionized gas 
recombination time. 
The resulting shell is ``momentum supported", rather than ``energy supported", 
and relatively thin compared to its scale.
Mass and momentum are conserved and transported within the shell. 
This momentum conserving assumption provides the possibility of
investigating the kinematics analytically. 
\citet{wil96} derived the formula for the shell's shape
by applying mass and momentum conservation
and neglecting the gas thermal pressure 
on both sides of the shell, as
well as the tangential acceleration caused 
by pressure gradients within the shell:
\begin{eqnarray}
r(\theta)&=&r_{0}\csc\theta\sqrt{3(1-\theta\cot\theta)},
\end{eqnarray}
where $r(0)=r_{0}$ is the standoff distance, 
and $\theta$ is the angle between the point on the shell
and the apex. He also derived expressions for the 
surface density and the tangential velocity of the shell.

\subsection{Relaxation}
The analytical method has no numerical uncertainties, 
but necessarily leaves out some details.
Once we
include more terms in the momentum equation, it can no longer 
be integrated analytically.
Assuming a bow shock is a steady state configuration, we can apply a relaxation
method to the mathematical description of the system. Using the relaxation
method allows us to include more physical processes than the analytical method
of \cite{wil96}, while being easier to formulate and less
time-consuming to calculate than a full hydrodynamic approach.

We assumed a cylindrically symmetric bow shock shaped as an approximately paraboloidal
shell. A grid with fixed
angular size is created on the surface of the shell and each cell
has the same angular size along both 
the azimuthal and the polar directions, $\delta\theta=\delta\phi=0.5^{\circ}$.
Due to the symmetry of such a geometry, 
we only need to calculate quantities in one strip of cells
along the polar direction.
We assume that the gas in a cell moves as a single fluid.
That is,
we assume rapid cooling behind the shock where the ambient and stellar
wind material mix and form a uniform shell, and we neglect shear motion
between front and back sides of a cell.
Mass flow, gas velocity, 
and position calculated from the analytical formula \citep{wil96}
are assigned to each cell as initial conditions (which were shown not to
affect the results).
Then, we let the program iteratively adjust these parameters by applying 
mass and momentum conservation.

During relaxation, the physical and geometric properties of each cell
are recalculated to replace the previous values in each step
until the whole system achieves convergence.
In each step, the mass flow from one cell (i) into a neighboring 
cell (i+1) is given by:
\begin{eqnarray}
\delta m^{i}=\delta m^{i-1}+\delta m_{w}^{i}+\delta m_{a}^{i} 
\end{eqnarray}
where $\delta m^{i-1}$ is the mass flowing out of the previous cell.
The mass flows from the stellar wind and the ambient medium 
are $\delta{m}_{w}^{i}$ and $\delta{m}_{a}^{i}$.
They are given by:
\begin{eqnarray}
\delta m_{w}^{i}&=&\frac{\delta\Omega}{4\pi}\dot{M}_{w}\tau \\
\delta m_{a}^{i}&=&-{\rho}_{a}({\mathbf{v}}_{a}\cdot{\mathbf{S}}^{i})\tau
\end{eqnarray}
$\delta\Omega$ is the solid angle subtended by the cell element,
$\dot{M}_{w}$ is the mass loss rate of the stellar wind,
${\mathbf{S}}^{i}$ is the outward pointing normal 
vector with magnitude equal the surface area of the cell $i$,
$\mathbf{v}_{a}$ is the velocity vector of the
ambient material, and $\tau$ is the time step of the iteration. 
Under the same
condition, we also have the formula for total momentum flow from the cell:
\begin{equation}
\delta{\mathbf{p}}^{i}=\delta {\mathbf{p}}^{i-1}+\delta m_{w}^{i}\mathbf{v}_{w}^{i}
+\delta m_{a}^{i}\mathbf{v}_{a}^{i}+(P_{w}^{i}-P_{a}^{i})\mathbf{S}^{i}\tau+\Phi_{P}
\end{equation}
where 
${P}_{a}^{i}$ and $P_{w}^{i}$ are the gas pressures of the external media.
The orientation of the
normal is perpendicular to the direction of tangential velocity,
${\mathbf{v}}^i = \delta{\mathbf{p}}^i / \delta m^i$.
$\Phi_{P}$ is the extra pressure term resulting from the 
tangential pressure gradient inside the shell.
We include it in the relaxation calculation after testing our model
for the case without pressure acceleration.
The distance of the cell from the star is calculated from the law of sines:
\begin{equation}
r^{i+1}=\frac{\sin(\beta^{i})}{\sin(\beta^{i}-\delta\theta)}r^{i},
\end{equation}
where $\delta\theta$ is the angular separation between cells
and $\beta$ is the angle formed by the radius and the velocity vector. 
Because we assumed that the system will eventually
reach a steady state, we did not try to solve a non-steady problem. 
Our calculation simply proceeds and finds a converging solution.

The shape of the shell and the tangential velocity along the shell,
calculated by the relaxation method neglecting the contribution 
of the gas pressure
gradient, are shown (dashed line) in Fig.~\ref{wwopressure}.
For this illustration, we set the star's speed equal to $20~$km~s$^{-1}$.
For comparison, the analytical solution \citep{wil96} is also shown
(dotted line). 
We also show an overall shape of the calculated shell in Fig.~\ref{shell}.
In this later calculation, we include gas thermal pressure contributions from
both sides of the shell by assuming a temperature of $50$K
for the ambient medium and $10^{4}$K for the free flowing stellar wind, 
which gives a more accurate but only slightly different solution
for the tail portion of the shell than the analytical solution.
We can see that the discrepancy between
our iterated result and the analytical solution \citep{wil96} 
is smaller before we add pressure 
acceleration to the relaxation solution. 
Calculations show that, in the rest frame of the star, 
gas starts to move with zero tangential 
speed from the apex of the paraboloidal like shell and accelerates until it reaches
the star's travelling speed, $20~$km{\,}s$^{-1}$, at the 
end of the shell. This acceleration is due to the accumulation of 
mass and momentum,
which are provided by swept-up ambient material and the stellar wind.
In later sections, we will show that this simple picture may not be
totally correct since the effect of the pressure gradient within the 
shell can be significant.

Fig.~\ref{wwopressure} also shows the surface density and the particle 
number density along the shell. The surface density increases, 
almost linearly, from 
its value at the apex of the shell.
The particle number density, derived from the shape of the shell and the normal
pressure components, drops rapidly
from its value at the apex.
It drops by a factor of four by the position where the shell passes the star. 
In steady state, gas in the shell should be in appoximate 
pressure equilibrium with 
the average of 
the normal components of the pressures on 
the two sides. 
The difference between the two normal components causes a pressure gradient
across the shell which causes its centripetal accelaration. From the pressure
in the ionized gas, we can calculate the 
density: $n=\frac{P_{ram\perp}+P_{gas}}{kT}$,
where we take the ionized shell's
temperature as $10^4$K.

\subsection{Pressure Gradient Acceleration}

\begin{figure}
\epsscale{0.7}
\plotone{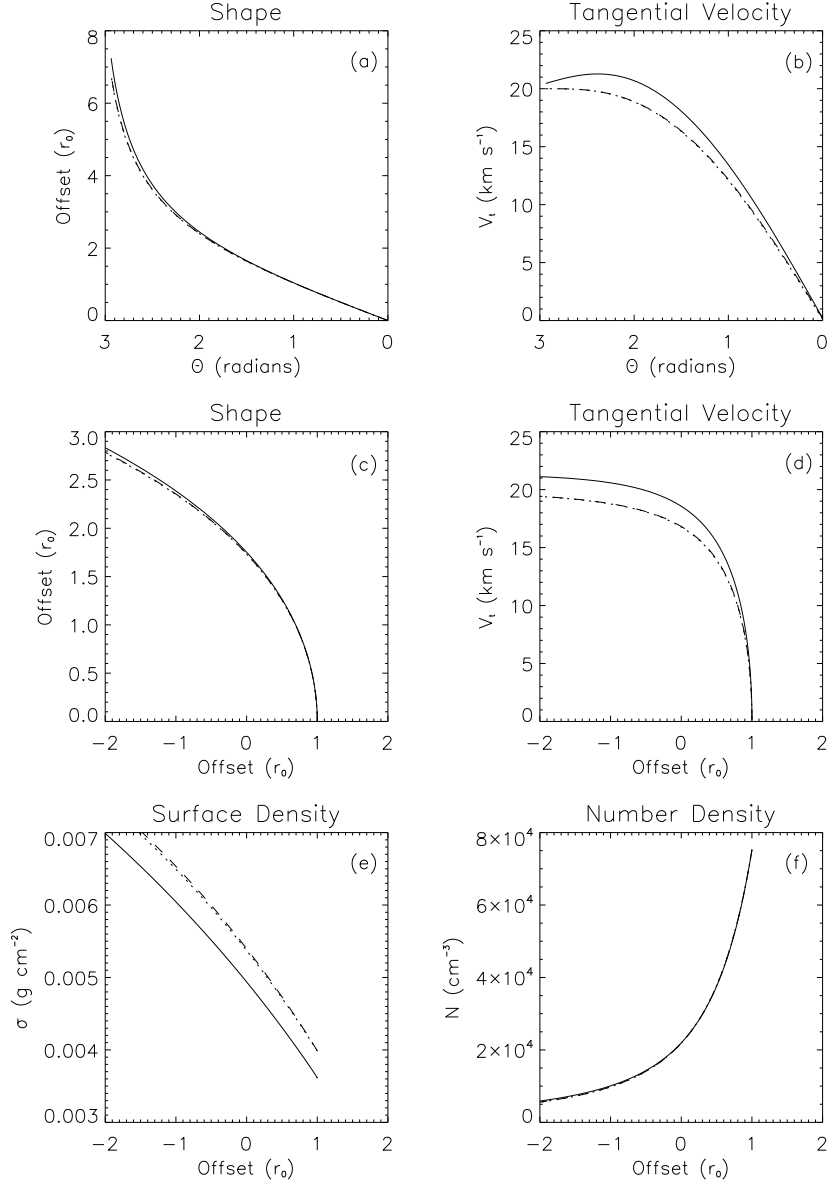}
\caption{Model for G29.96~-0.02.
(a) The shape of the shock, plotted against the angle from the apex ($\Theta$).
(b) The tangential velocity $V_t$ along the surface.
(c) The shape of the shock. X is the symmetry axis of the shell.
Offsets are measured from the position of the ionizing star. 
The length unit is the standoff distance ($r_0$).
(d) The tangential velocity $V_t$ along the shock surface.
(e) Surface density ($\sigma$).
(f)The particle number density N along the bow shock.
The results from \cite{wil96} are shown with dotted lines.
The model without the pressure gradient is shown with dashed lines.
The model with the pressure gradient is shown with solid lines.
Notice that the results of the model without the pressure gradient is extremly close
to the results of \cite{wil96}.
\label{wwopressure}}
\end{figure}

\begin{figure}
\epsscale{1.0}
\plotone{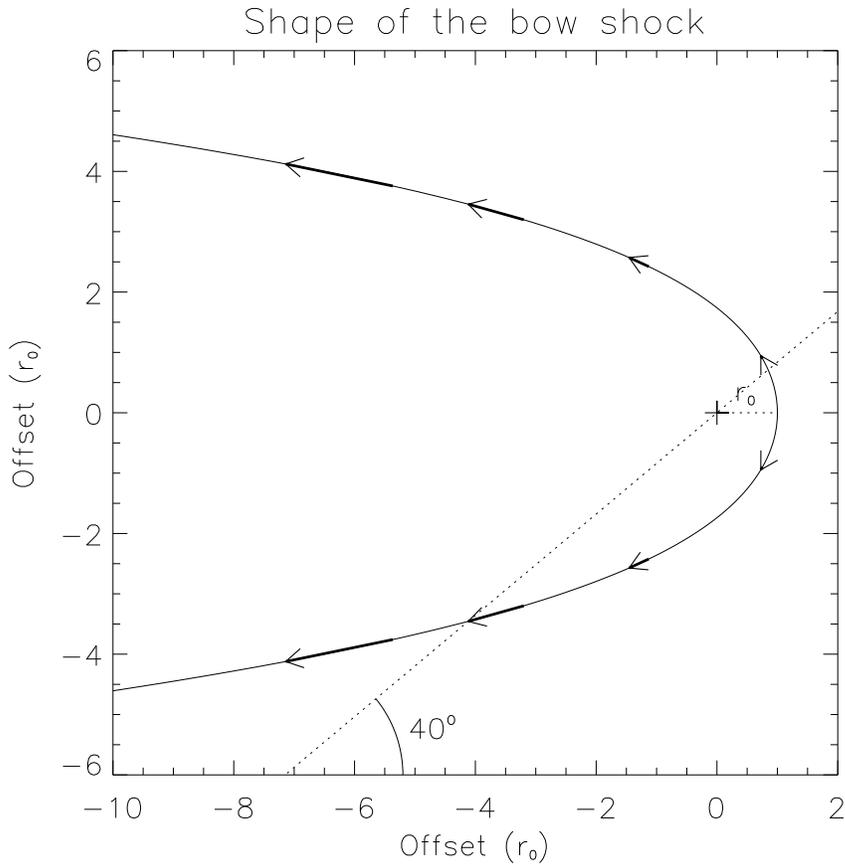}
\caption{The calculated shape of the bow shock without the pressure gradient 
contribution and the corresponding 
tangential velocity vectors in G29.96~-0.02 using parameters of model~1
in Table \ref{partab}. 
The cross at (0,~0) indicates the position of the star. $r_{0}$, the standoff distance from the
star to the apex of the shell, is the length unit in the plot. The dotted line indicates that the
line of sight from the lower left to the upper right, form $40^{\circ}$ with the symmetry axis of
the shell.
\label{shell}}
\end{figure}

Most existing solutions, 
both analytical \citep{wil96} and numerical \citep{macVW91}, 
neglect the effect of the gas pressure gradient along the compressed shell.
The exception is \citet{comK98}, who make a numerical
hydrodynamic calculation, but concentrate on the case of a runaway O star
moving through the diffuse ISM.
In addition to the centripetal acceleration 
caused by the unbalanced external pressures, 
the variation in the pressure along the shell results in
a pressure gradient that accelerates the gas along the shell.
The momentum deposited in a cell by this effect is given by:
\begin{eqnarray}
\Phi_{P} = \frac{P^{i-1}-P^{i+1}}{2} A^{i} \tau
= {\mathbf\bigtriangledown}{P}^{i} V^{i} \tau
\end{eqnarray}
Where $P^{i}$ and ${\mathbf\bigtriangledown}{P}^{i}$
are the pressure and the pressure gradient in the $i$th cell.
$A^{i}$ and $V^{i}$ are the cross section and the volume of the cell.
We derive the thickness of each cell from the emission measure, using the 
ionization-recombination equilibrium equation, and the density, 
calculated from the
balance between gas pressure and ram pressure assuming stellar parameters
appropriate for G29.96 (section 4.4.1).
We then assume that the thickness of the given cell is constant.
If the central star is unable to ionize the whole shell,
the thickness of the shell is slightly greater than that
of the ionized layer, with
the neutral gas in a thin, dense region just outside of the HII region. 
The thickness increases from the apex of the shell to the tail region
because the density drops along the shell. 
The results of the relaxation are shown with a solid line in Fig.~\ref{wwopressure}.
The analytical solution is shown as a dotted line for comparison.

From the plots in Fig.~\ref{wwopressure}, 
we can see that the change to the shape of the shell 
caused by including the pressure gradient is very small.
Thus the change in the ionized gas number density is also small, because
it is calculated from the ram pressure normal component, which
directly relates to the shape of the shell. In the region close to the apex,
this change can be neglected.
A larger change is seen in the tangential velocity plot.
For the parameters assumed, which includes a partially ionized
shell, the velocity near the head of the bow shock increases so that
the gas speed is $\sim 1~$km~s$^{-1}$ higher than without the pressure
gradient acceleration at $\theta\sim1~radian$.
Without the pressure gradient,
the maximum velocity that gas can achieve is the speed of the star.
With the additional acceleration provided by the pressure gradient,
the gas in the shell can reach higher velocities.
The tangential speed finally drops 
as the shell picks up more mass from the ambient medium and the
pressure gradient acceleration effect decreases
toward the end of the shell due to the drop of stellar wind ram pressure.
With the parameters we have chosen, the maximum speed that gas 
can achieve under the
pressure gradient acceleration is $1.3~$km~s$^{-1}$ higher than 
the speed of the star.
Up to this point,
we have assumed that the ionized gas and the swept-up neutral gas
are coupled, so they share the momentum carried by the stellar
wind and ambient medium.
When including the effect of a pressure gradient, we assume that
the force resulting from this effect 
is also shared by ionized and neutral gases.
In fact, most of this force should be exerted on the ionized gas only, 
since a pressure
gradient causes a force per unit volume and the ionized gas fills most of
the volume.
If the ionized gas can slip past the neutral layer, it should reach a higher
velocity than what we present in this work. 
The resulting speed should be similar to that in
a fully ionized shell situation.

If the optical
depth of the shell is small for ionizing radiation, the ionizing photons
will ionize the whole shell or penetrate the shell and ionize 
the ambient medium beyond it.
In this ``ionizing photon
leaking'' situation, the pressure gradient acceleration effect will
reach its maximum because the force is proportional to the thickness 
of the ionized shell.
Our calculations show that 
the gas speed in the shell can reach
1.25 times the star speed in the range 10~-~20~km$~s^{-1}$,
and there is no drop in speed for angles larger than 2.4 radians
seen in Fig.~\ref{wwopressure}(b). 

In this work, we only show
the results of the single layer bow shock model. 
Including multiple 
layers to our model is a logical next step.
Furthermore, we leave out the pressure gradient in the 
hot post-shock gas on the inner side of
the shell from our model. Since it is so hot (${\sim}10^7$~K),
it should cool very slowly. It is
even harder to include the contribution from this part of HII regions
in the model.

The pressure gradient accelerates gas inside the shell,
thus decreasing the surface density. It does not affect our flux maps
because the brightness of each cell in our model depends only on the number of 
ionizing photons, as long as the shell is only partially ionized.
However, the change in the surface density will affect
the thickness of the neutral part of the shell, thus affecting the
appearance of corresponding molecular flux maps. A
kinematic study of the neutral component associated
with ionized bow shock structures might prove useful.

If pressure gradient acceleration is ignored,
the spatial and kinematic structure of a modeled bow shock depend
on two parameters: the speed of the star and the stand-off
distance.
The dependence of the stand-off distance on the stellar mass-loss rate,
stellar wind terminal speed, stellar speed, and the ambient medium
density is given in Eq.~2.
As is discussed above, inclusion of the effect of the pressure gradient in
the ionized gas has little effect on the spatial distribution, but changes
the velocities much like an increase of the stellar speed would.
A set of models for UCHII regions in G29.96 and Mon~R2 is given in Table
\ref{partab}, where plausible values, based on other observations 
(see references in the table), are given
for unconstrained parameters.

\section{Comparison with Observations}
Our relaxation models of bow shocks
provide the shape of the shell and the Doppler shift at each
point on the shell.
We calculated the [Ne~II] emission at each point on the shell,
assuming
ionization-recombination balance in the shell, taking into account the angle
at which the ionizing radiation hits the shell. 
We ignore any dust that might
be present in the region.

The model ionized shell was tilted relative to the line of sight to improve
the agreement with the data, sampled and remapped onto the sky with
a spatial grid $0.2''\times0.2''$, and then convolved with TEXES's
velocity resolution of $4~$km~s$^{-1}$
and a model of TEXES's beam of half-maximum radius $r\sim0.8''$.  
For G29.96, we convolved with an additional turbulent linewidth of
$10.0~$km~s$^{-1}$ to improve the agreement with the observations.
In Mon~R2, we convolved the data with the the narrowest line 
width ${\sim}8.8~$km~s$^{-1}$ \citep{jafZLR03}.
Our observing sampling grid is much coarser than our modeling grids
in the region corresponding to our observations, 
so the error caused by the quantization of the modeling grids is small. 

\subsection{G29.96~-0.02}
Four model parameters are constrained with varying 
uncertainties by our observations:
the standoff distance ($r_0$), the tilt of the 
symmetry axis from our line-of-sight,
the speed of the star relative to the ambient 
medium and LSR velocity of the 
star (or equivalently of the molecular cloud). 
In addition, the stellar ionizing luminosity
and spectral type, and the neon abundance combine 
to determine the [Ne~II] brightness.
Other parameters, notably those from which the 
standoff distance can be calculated, can
be determined from other observations of our 
objects or estimated from typical O-star
properties. Parameters determined in this way are given in Table \ref{partab}.

We rotate the observed flux 
datacube $26^{\circ}$ \citep[assuming p.a.=$64^{\circ}$,][]{marBKTH03} 
counter-clockwise in order to  
orient the axis of the shell horizontally in the figures. 
We also position the observed flux map so that the position 
of the ionizing star matches that in our model.
A cross at (0,~0) in the map (Fig. \ref{intmap}) 
indicates the position of the ionizing star. 
The spatial distribution of the [Ne~II] emission is reproduced better if
the motion of the star in the model is within $45^{\circ}$ of the plane of
the sky, but a bigger angle tends to give a better fit 
in position-velocity diagrams.
We chose to tilt the shell $50^{\circ}$ away from us in our model.

As discussed in section 4.3, we assume that the ionized gas 
and the neutral gas in the shell are moving together, and we pick
parameters so that the shell is only partially ionized.
A stellar speed of $v_* \sim\,20\,$km\,s$^{-1}$ is picked
to fit the observed range of velocities. An LSR velocity
of the star of $V_{*,LSR}=104~$km\,s$^{-1}$, or $V_{amb,LSR}=89~$km\,s$^{-1}$
is needed to fit the velocity offset observed. However, observations
of the nearby molecular 
material \citep{chuWC90,cesWC92,affchk94,olmC99,lumH99,marBKTH03}
give $V_{amb,LSR}=92-100~$km\,s$^{-1}$, 
with most numbers near $98~$km\,s$^{-1}$.
A stellar speed of $v_* \sim\,10\,$km\,s$^{-1}$ would 
allow better agreement between
the model velocity offset and the molecular observations, but would not fit
the observed [Ne~II] velocity range unless the 
pressure gradient acceleration is larger
than in our model.

\begin{figure}
\epsscale{1.00}
\plotone{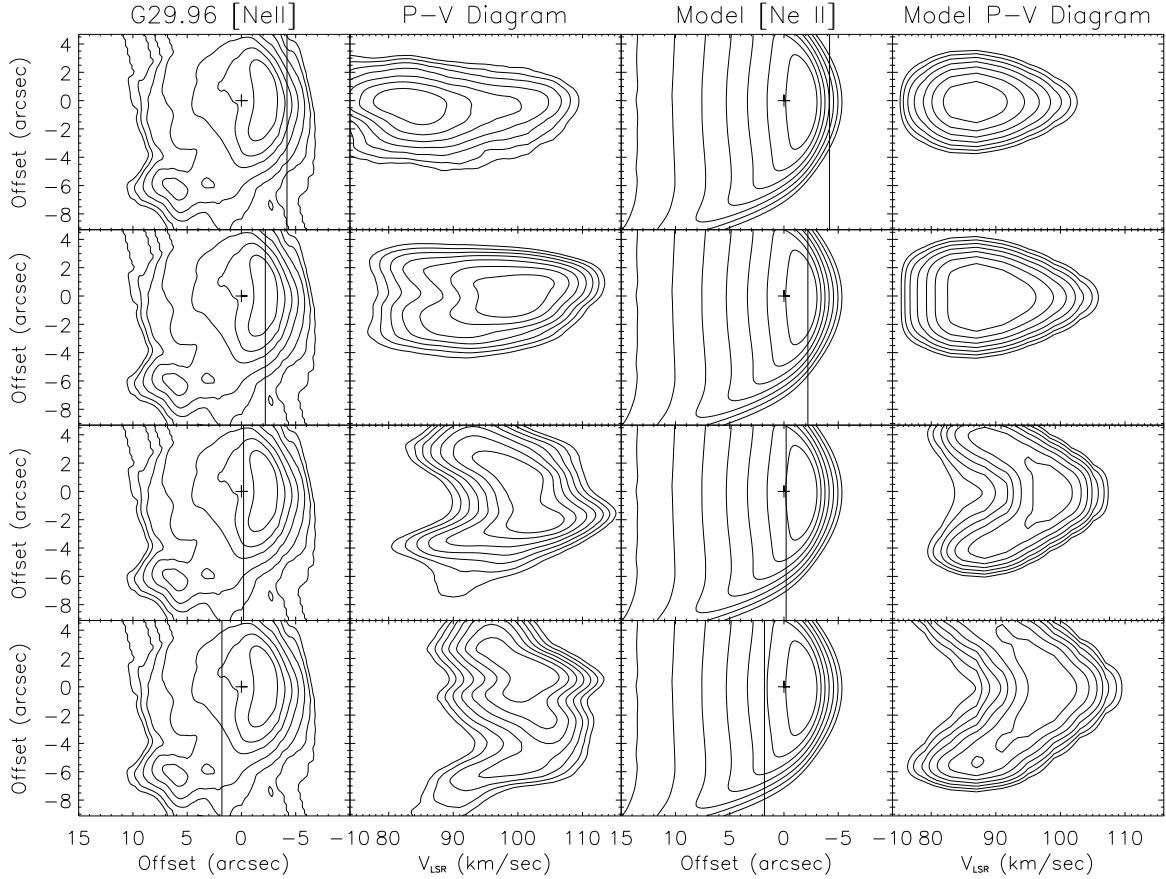}
\caption{Position-Velocity (P-V) diagram of 
G29.96~-0.02 [Ne~II] observation and that of model~1 (Table \ref{partab}).
Contours in the flux map are drawn at $50\%$, $25\%$, $12.5\%$,
$6.25\%$, $3.12\%$, $1.56\%$ and $0.78\%$ of
the peak value. Contours in the P-V diagrams are drawn in the similar way, 
but with contour levels separated by $\sqrt{2}$.
A cross at (0,~0) indicates the location of the star according to the
model. Same for the following plots.
\label{g29step}}
\end{figure}

\begin{figure}
\epsscale{0.5}
\plotone{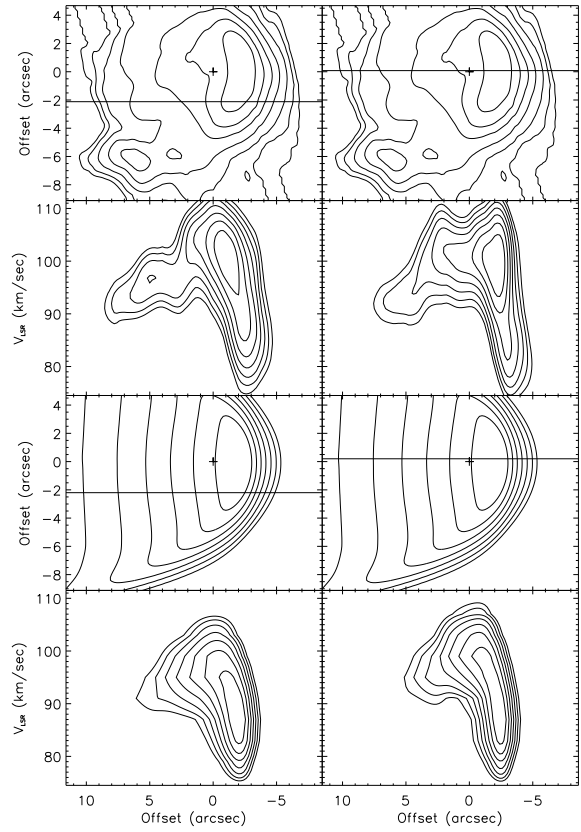}
\caption{Position-velocity diagram of G29.96~-0.02 [Ne~II] observation (top panel) and model~1 (bottom panel).
\label{g29pix}}
\end{figure}

The simulated [Ne~II] flux map and P-V diagrams on cuts 
through various positions in the model 
are shown with corresponding diagrams from observed data in 
Fig.~\ref{g29step}-\ref{g29pix}s.
The accelerating, paraboloidal like flow produced in the bow shock model successfully 
matches many global features in the observed data.
The model produces the limb brightening at the head region,
although the fit would be improved by using a smaller tilt away from the
observer.
The curvature of the bright ridge is also fit well.
Cuts perpendicular to the shell axis show similar central velocity shifts,
spatial and spectral ranges, and overall shape, including a ``$>$'', 
in both the observations and the model (Fig.~\ref{g29step}).
From cut to cut, the curvature of the ``$>$'' changes 
because the motion of the shell with the star gives more 
redshift to gas closer to the apex than to down-stream gas. 
Our line-of-sight passes through the ionized shell twice for
positions near the symmetry axis. Given the orientation of the UCHII region,
the far side is closer to the apex, contains higher density gas, and dominates
the line emission. The near side is farther from the apex and has lower density
gas. At the edges of the cuts, we also see more lower density gas.
The ``$>$'' forms because gas closer to the apex is less blue-shifted, compared with gas
farther from the apex.
Because the cuts are perpendicular to the symmetry axis, the model
P-V diagrams are necessarily symmetric. 
Asymmetries in the observations are also small.
In cuts parallel to the symmetry axis (Fig.~\ref{g29pix}), 
the P-V diagrams show a ``7'' like pattern, with a curved 
leg and a rather flat top ``arm'', 
indicating a large velocity gradient at the head 
and less velocity change towards the tail.
The line is also broader at the head region. These are
predicted by the model too. In our velocity plot (Fig.~\ref{wwopressure}), 
we can see that more than $80\%$ of the change in the 
tangential velocity happens in the first $\pi/2~radians$. 
The line is broader because the scale length for velocity 
change is smaller in the head region. The remarkable similarity
of the models and the observations is a clear indication for 
a large-scale paraboloidal like flow in this UCHII region.

\subsection{Mon~R2 IRS1}
\begin{figure}
\epsscale{0.8}
\plotone{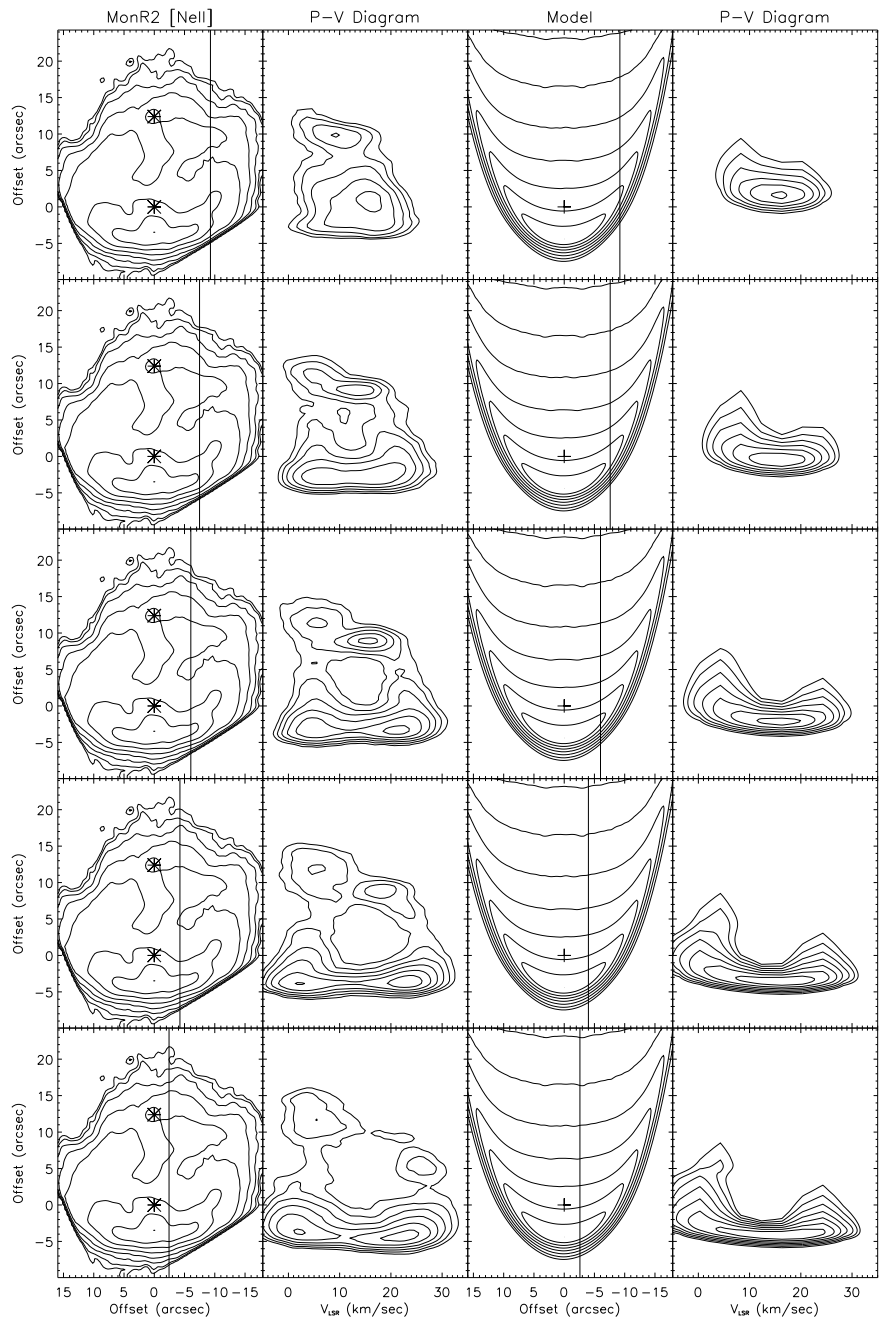}
\caption{
UCHII region in Monoceros~R2 and model~2 with the standoff distance matching that in \cite{yaohinoswy97} (Table \ref{partab}). [Ne~II] map and the
position-velocity diagrams along the corresponding cutting lines.
Two asterisks and one cross mark the position of IRS1 (0,~0) and IRS2 (0,~12.4) from \cite{yaohinoswy97} and
the position of the ionizing star (0,~0) used in the model.
\label{monstep1s}}
\end{figure}

\begin{figure}
\epsscale{0.8}
\plotone{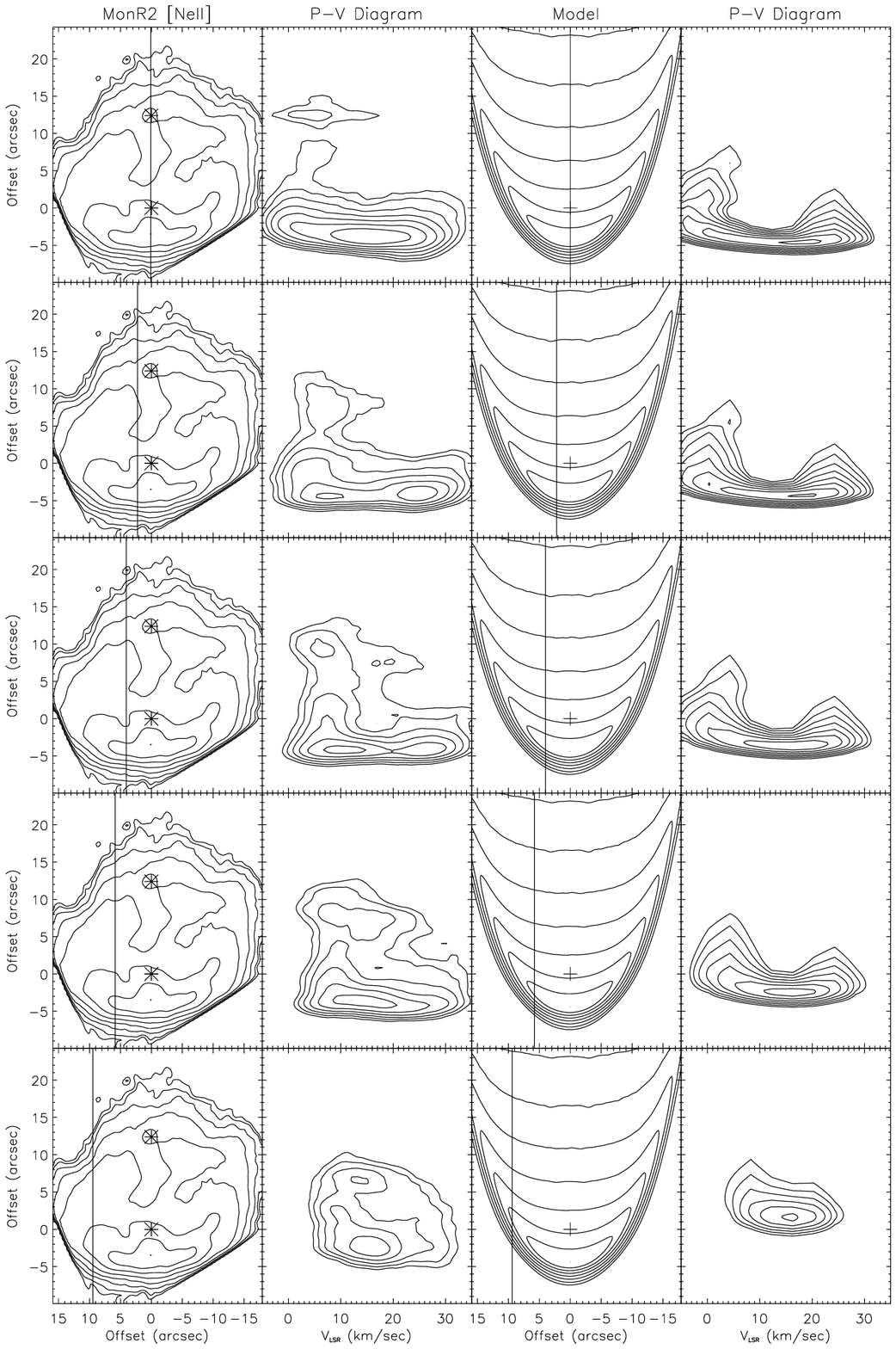}
\caption{Same as in Fig.~\ref{monstep1s}.
\label{monstep3s}}
\end{figure}

\begin{figure}
\epsscale{0.8}
\plotone{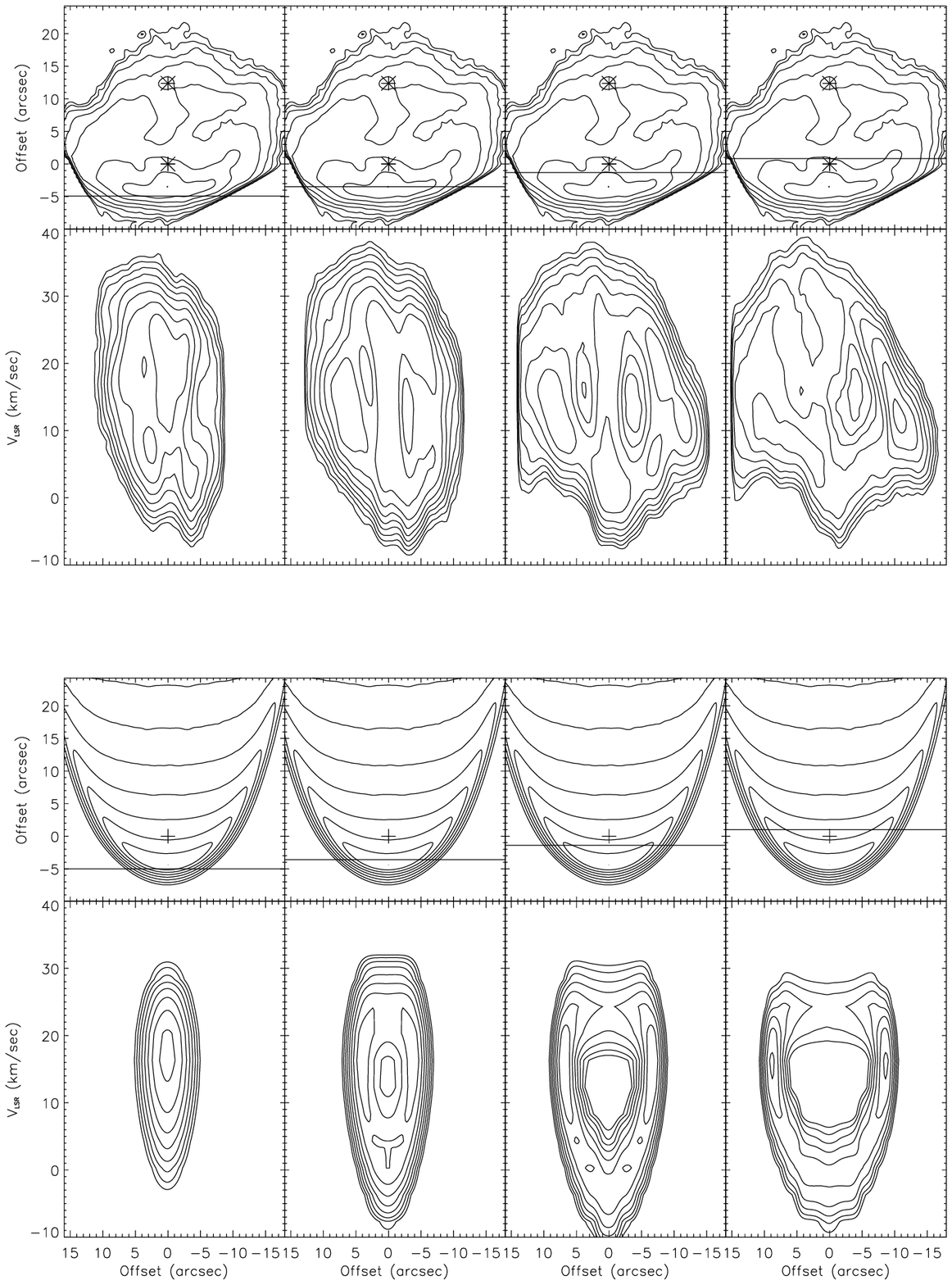}
\caption{Same as in Fig.~\ref{monstep1s}.
\label{monslit1s}}
\end{figure}

\begin{figure}
\epsscale{0.8}
\plotone{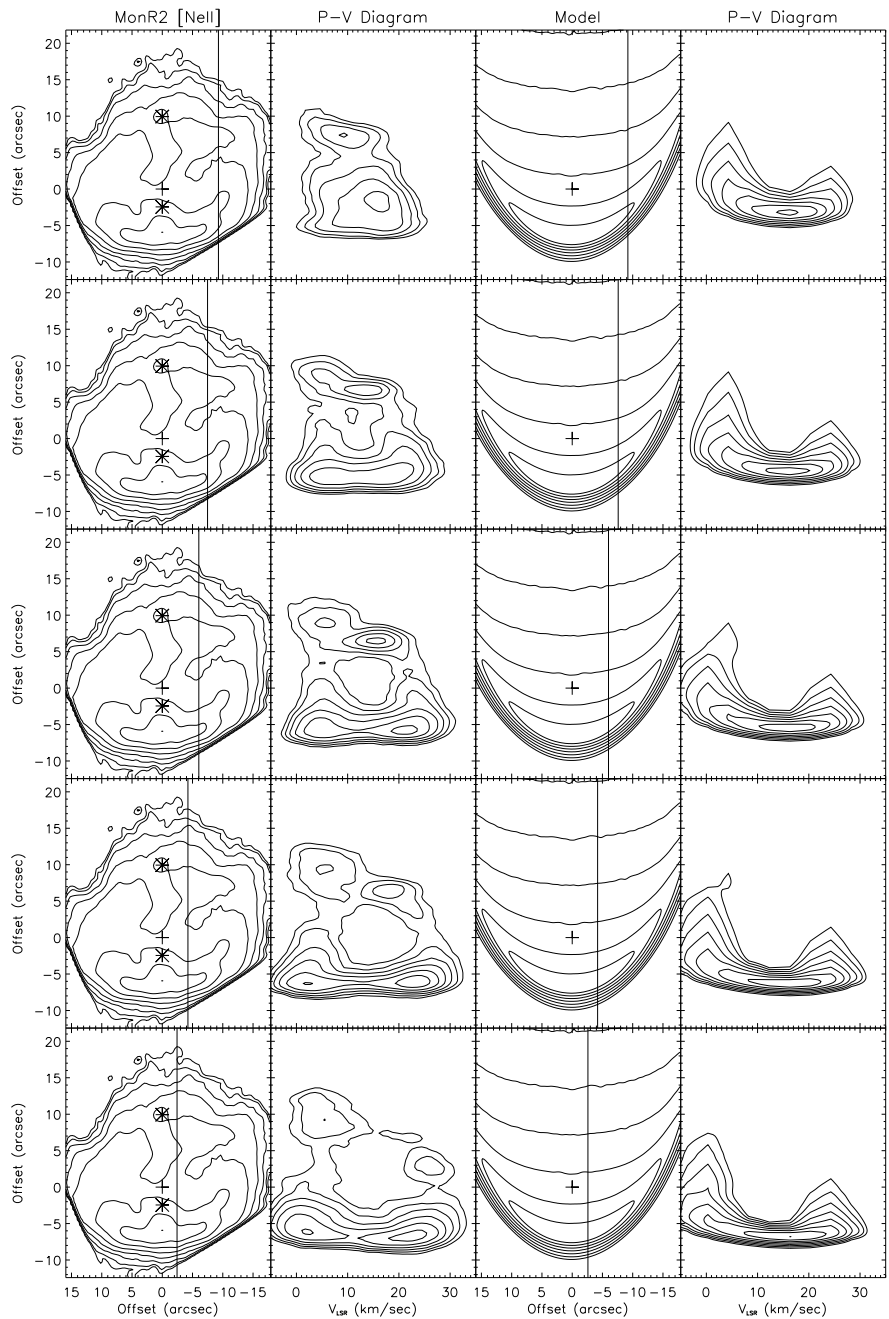}
\caption{
HII region in Monoceros~R2 and model~3 with a longer standoff distance than 
that in \cite{yaohinoswy97} (Table \ref{partab}). [Ne~II] map and the 
position-velocity diagrams along the corresponding cutting lines.
Two asterisks and one cross mark the position of IRS1 (0,~-2.45) and IRS2 (0,~14.8) from \cite{yaohinoswy97} and 
the position of an ionizing star (0,~0) according to the model. 
\label{monstep1}}
\end{figure}

\begin{figure}
\epsscale{0.8}
\plotone{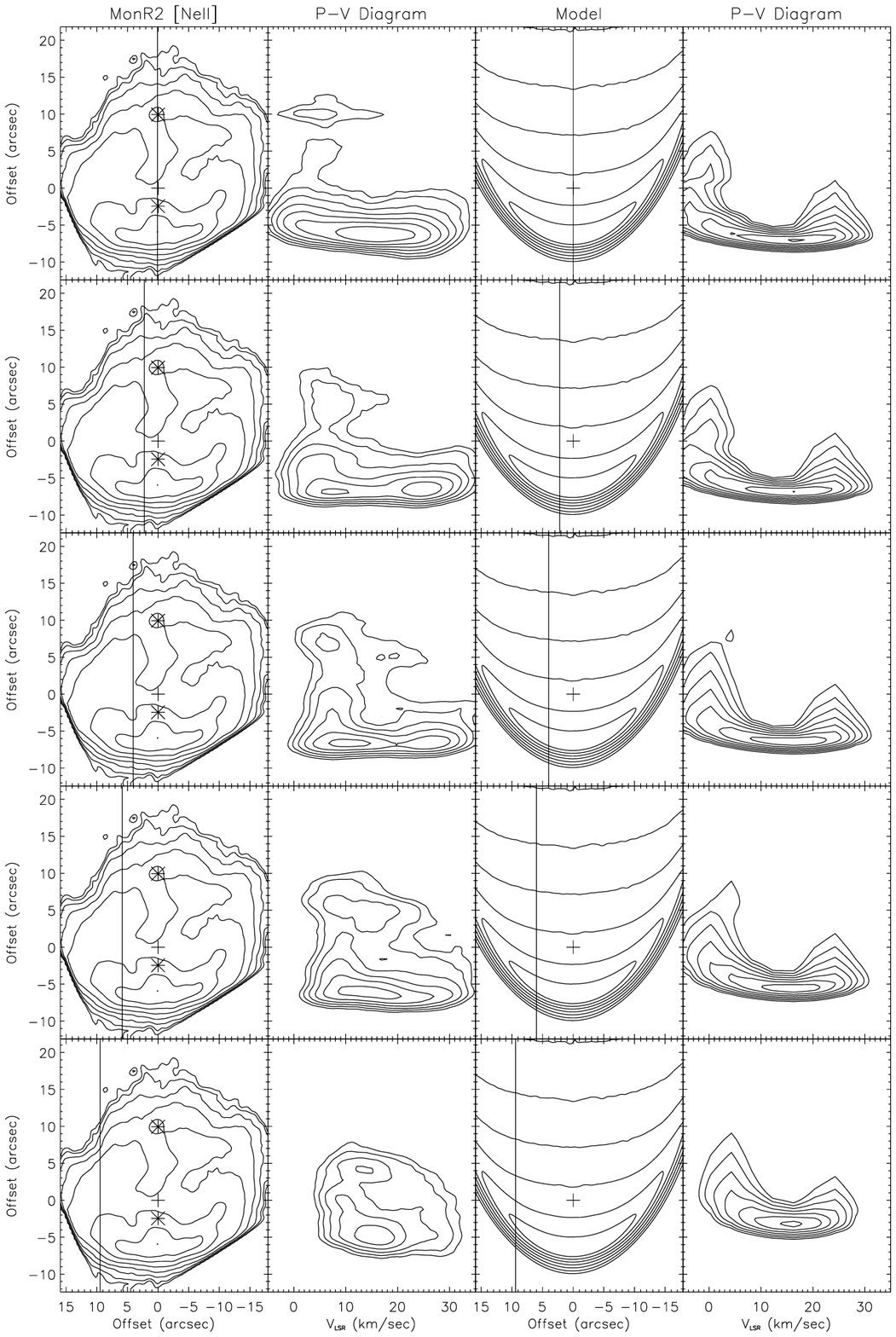}
\caption{Same as in Fig.~\ref{monstep1}.
\label{monstep3}}
\end{figure}

\begin{figure}
\epsscale{0.8}
\plotone{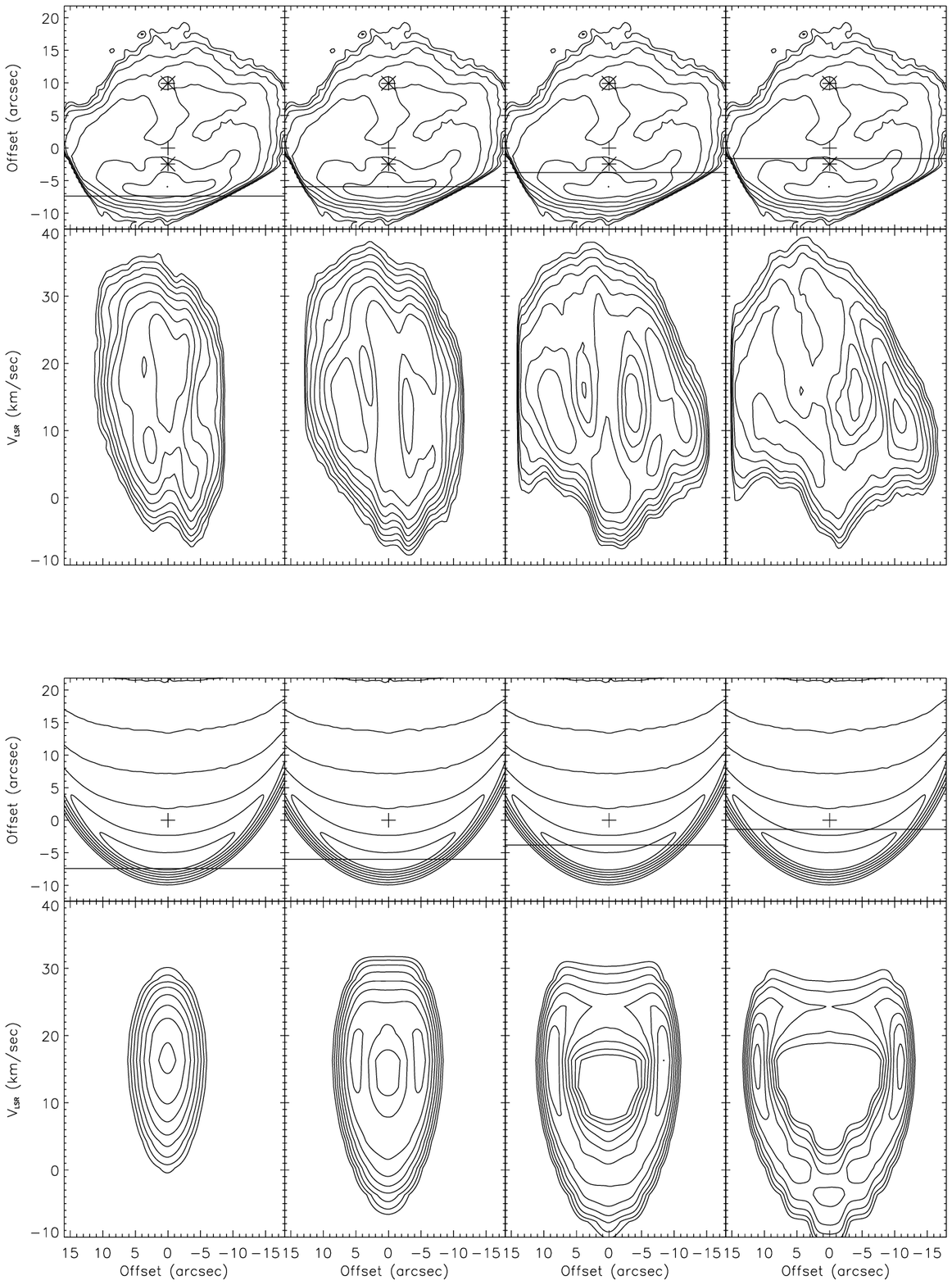}
\caption{Same as in Fig.~\ref{monstep1}.
\label{monslit1}}
\end{figure}

Our observations of the compact HII region in Mon~R2~IRS1 are presented in 
a previous paper \citep{jafZLR03}. 
With a $24''$ diameter shell and a bright southeast ridge,
[Ne~II] emission in Mon~R2 shows complex and broad velocity structure.
We speculated that we were looking at 
a kinematic pattern in which material flows from 
the bottom to the rim of a bowl-like feature. 
Here, we try using the bow shock model to interpret the observations,
since the object also has a cometary appearance. 
In order to 
show the similarities between the kinematics of a shell-like flow
structure
and the observed data,
we choose parameters so that
the pressure gradient acceleration is negligible.
The results of two models with different standoff distances are shown
in Figs.~\ref{monstep1s}-\ref{monslit1}. In the first model, we try to match
the ionizing star position to that in 
near-infrared observations that are good to $<1''$ \citep{yaohinoswy97}. 
In the second model (Table \ref{partab}), 
we use a bigger standoff distance, which
gives a better fit to our [Ne~II] observations.
In both cases, we tilt the shell $20^{\circ}$ toward 
the observer in the simulation
for a good fit to the observed P-V diagrams. As in the G29.96~-0.02 case, 
we derive a $V_{amb,LSR}$ different from the value we 
found in the literature for Mon~R2. 
An additional $\sim8$~km~s$^{-1}$ redshift is 
needed to shift the center of the line to the value
in the observational data. 
It is interesting to note that Mon~R2 requires a redshift and is
tilted toward us while G29.96 is tilted away and requires
additional blueshift. This may indicate the pressure gradient
may have a bigger effect on the gas acceleration than we
calculate and that a more sophisticated model might fit the data
with a smaller stellar velocity.

Qualitatively, the model agrees with the data.
The observed morphology appears cometary, with a bright arc and a fainter
tail, although the tail ends more abruptly than in the model.
The P-V diagrams show two peaks over most of the region, as is expected where
our line of sight passes through 
the front and back sides of a shell. 

If the predominant motion were due to expansion of the shell, rather than
flow along the surface, the velocity splitting would increase only gradually
moving from the edge of the shell into the center.
As in G29.96, the motions in Mon~R2 are consistent with gas flow along a shell.
However, there are a few facts which make Mon~R2 HII region hard to fit into
the bow shock model. 
First, there is a compact broad-lined region near or just inside of the apex
of the observed shell, which is not predicted by the model.
It is most apparent in Fig.~\ref{monslit1s} and 
Fig.~\ref{monslit1}, where it produces the central
ridge in the P-V diagrams. This component is also shown 
in our previous paper \citep[Fig.~7]{jafZLR03}.
This source may be a result of the shock front overtaking a
dense clump in the molecular cloud,
or, it could result from an instability in the front, as is
seen in the hydrodynamic models of \citet{comK98}.
The second difficulty is that a larger standoff distance is required
to make the curvature of the shell the same as in the data.
In the first cut of Figs.~\ref{monstep1s} and~\ref{monstep1}, where
our observed P-V diagrams
only show one component, our model P-V diagrams show two components.
In addition, the shell appears to be closed on the back side.
The outermost contours on the images are nearly circular,
although the rim is much brighter at the bottom of 
the map than at the top. The model
predicts that the shell is always open toward the tail.

In our figures~\ref{monstep1s},\ref{monstep3s} 
for both the model and the data, 
cuts parallel to and near the shell axis show two components.
At the ridge of the shell-like region,
these two components are connected by a broad line 
with a width up to $40~$km{\,}s$^{-1}$. 
The relative strengths and the separation of the two components
changes with the position of cuts. Normally, the blue-shifted component 
is seen farther toward the tail because the tilt of the shell 
puts the denser part of the near side of the shell farther
from the head than the far side. 
The single broad line in the P-V diagram of the cut made in the center
of the ridge is probably caused by the additional source there. 

The observed P-V diagrams of the 
cuts perpendicular to the shell axis (Fig.~\ref{monslit1s},\ref{monslit1})
demonstrate the gradual change of a circular structure,
which is typical for cuts across a rotationally symmetric shell.
If we neglect the broad-line component at the peak of the flux map, 
our model and observed P-V diagrams agree well.
Compared to the observations, the model with a smaller standoff distance 
predicts a smaller spatial span (Fig.~\ref{monslit1s}).
Once we increase the standoff distance (Fig.~\ref{monslit1}), 
we no longer have this problem.
A density gradient along the symmetry axis of the shell
might be able to explain the curvature of the shell,
although there is at present no concrete evidence for such a gradient.

\section{Discussion}

We have shown that the overall velocity structure 
behind the cometary shape of UCHII
regions in G29.96~-0.02 and Mon~R2 is formed when
ionized gas flows along a paraboloidal like surface. 
Using the bow-shock model, we can reproduce this 
kind of structure. We found that 
the model
provides a good qualitative explanation of our
observations of G29.96 and Mon~R2.
Both sources have morphologies similar to that predicted by the model,
at least near the apex of the shell.
The observed line profiles, as seen in the P-V diagrams, are generally
double-peaked, indicating that the gas is swept up into a thin shell, so
each line of sight passes through two surfaces with different Doppler
shifts.
The two Doppler components reach a maximum separation near the shell
apex, indicating that the gas accelerates along the shell, and that the
dominant motion is along the shell rather than radial.

There are quantitative differences between the model and the data,
especially in the case of Mon~R2, as discussed above, indicating that a
complete description of these sources will have to be more complicated than
our simple model.
Nevertheless, we think that the bow shock model must be essentially valid.
Now we consider whether this model resolves any of the problems associated with 
the study of UCHII
regions, and how the model might be improved.

The current bow shock model assumes that the stellar wind material 
and the ambient medium mix well and cool
efficiently after they collide, so their mass and momentum
contribute to the swept-up layer and the velocity of the ionized
gas is that of the single swept-up layer. 
However, this momentum-conserving 
thin shell assumption is not well justified.  
Behind the shock, stellar wind material is collisionally
heated and ionized. This gas will have a temperature  $>10^7{\,}$K and
does not cool efficiently, keeping the layer thick.
This high temperature layer forms between the
stellar wind and the photonionized gas layer and makes the mass and momentum
exchange between the two layers difficult.
The thin shell assumption is invalid in this situation. 
The photonionized layer is also evaporated through conduction with
the extremly hot shocked stellar wind gas, which makes
predicting the kinematics more difficult.
There are many instabilities which 
can inflate the shell to make the thin shell assumption invalid.
Possible candidates 
include the transverse acceleration instability (TAI) \citep{dgaVN96}
caused by the acceleration of the flow normal 
to the surface of the shell and the non-linear thin shell instability (NTSI) 
\citep{vis94, hue03}
caused directly by the collision of isothermal flows. 
These instabilities will disturb the
shell and create sub-structures with scales 
comparable to the thickness of the bow shock. Under such conditions, the
momentum conservation assumption will not hold.

Among the other models proposed to explain cometary UCHII regions, 
the champagne model is the most interesting.
The model 
applies when a massive star is found
in a region with a large density gradient, such as 
the edge of a dense molecular cloud.
The resulting HII region will
expand supersonically away from the high-density region, or simply break out of 
the edge of the molecular cloud
and cause the ionized gas to stream out of the opening 
in response to 
a large pressure gradient. 
The classic champagne model without a stellar 
wind \citep{ten79, bodTY79,yorTB83}
can explain the cometary shape of HII regions but, because
the ionized gas fills up the bubble and the main pressure 
gradient is along the density gradient, 
the champagne model has difficulty 
accounting for limb brightening and for the line profiles observed.
Adding a stellar wind to the champagne model will probably produce
limb brightening \citep{com97}.
Without ram pressure of the external gas, the champagne model with a stellar
wind tends to produce
a bigger shell in the clouds with the same density and temperature as the clouds
we investigated with the bow shock model.

The main differences between 
the bow shock and the champagne flow are seen in 
the kinematic properties of the ionized gas \citep{garL99}. 
First, the bow shock model predicts that the velocity
gradient is steeper in the head than in the tail, 
which can be seen easily in our P-V diagrams, 
whereas the 
champagne model expects the largest gradient
in the tail, where the fractional pressure gradient 
also reaches the highest value \citep{com97}. 
Second, the champagne model predicts that the line widths 
are broader in the tail region because the gas has higher
speed there and the gas motion is less parallel,
while the bow shock predicts broader line widths near the 
apex because the gas gains more
momentum there
and moves cylindrical-symmetrically along the surface 
(in the frame of reference of the star). 
Finally, ionized gas near the apex of the cometary 
structure is at rest with respect to the ambient molecular 
gas in the champagne model, whereas it moves with 
the star and/or the shock front in the bow 
shock model. In all of these respects, our observations 
of G29.96 and Mon~R2 are in better agreement
with the bow shock model. 

A central problem about UCHII regions is their lifetimes. 
The number ratio of UCHII
regions and OB stars
requires that UCHII regions should have an average lifetime $\sim~20\%$ of
that of OB stars \citep{wooC89a}. Any model of UCHII regions has to be able to
account for this. 
We have observed over a dozen of UCHII regions. They all
have large velocity range, usually over $\sim20$~km~s$^{-1}$.
In the bow shock model, the velocity range of the line is 
directly connected to the star speed
relative to the ambient medium. 
If $20\%$ of UCHII regions have a bow
shock-like structure, as shown in \cite{wooC89a, wooC89b} and \cite{kurCW94}, 
the same percentage of OB stars 
should move supersonically through molecular clouds. 
Although evolved OB stars can be accelerated up to $200 $km{\,}s$^{-1}$ through
the association ejection or supernova explosions \citep{bla93} 
and form bow shock-like
structures around them \citep{vanND95,norVD97,kapVAGPWZ97}, 
high speed OB stars are rare in OB associations.
The velocity dispersion of OB association is generally small, 
only a few km~s$^{-1}$ \citep{jonW88,tiaVZS96}.
Even if the molecular gas initially made the potential a bit deeper,
stellar speeds of $15-20$~km~s$^{-1}$ are extremely improbable
as part of a normal distribution.
Including pressure gradient acceleration 
helps to reduce this requirement. Our calculations suggest that
the needed speed is still too
high to solve the problem. This disagreement indicates that the stellar speeds
may be less than in our model. Due to the simplicity, our model could 
underestimate the acceleration of the ionized gas along the shock front. 
A full hydrodynamical treatment would help determine the true effect of
a pressure gradient.

\acknowledgments

We are grateful to Alan Fey and Ed Churchwell for letting us use their
VLA 2~cm data \citep{feyGCV95,wooC89b}. We thank Gregory Shields 
for help on the ionization model.
We also need to thank NASA IRTF staff for their help on observations.
This work was supported by NSF grant AST-0205518, NSF grant AST-0307497
and NASA grant NNG04GG92G. Thomas Greathouse is currently supported 
by the Lunar and Planetary
Institute, which is operated by the Universities Space Research
Association under NASA CAN-NCC5-679.

\clearpage

\clearpage

\clearpage

\clearpage

\begin{deluxetable}{ccccccccccc}
\tabletypesize{\scriptsize}
\rotate
\tablewidth{0pt}
\tablecaption{Bow Shock Model parameters \label{partab}}
\tablehead{
Object & Spectral& Object& Cloud& Ambient& Stellar& Stellar& Stellar& Standoff&Size On&Tilt\\
Name& Type& Distance& Radial&Medium&Mass loss&Wind&Travelling&Distance&The Sky&Angle\\
& & & Velocity&Density& Rate&Speed&speed&&\\
& & D & V$_{LSR}$ & $n_{a}$ & $\dot{M}_{w}$ & $v_{w}$ & $v_{a}$ & $r_{0}$&& \\
& & (kpc) &(km{\,}s$^{-1}$) & ($10^{5}$cm$^{-3}$) &($10^{-6}~M_{\odot}~$yr$^{-1}$) & (km{\,}s$^{-1}$) &(km{\,}s$^{-1}$) & ($10^{16}$cm)&$ $&
}
\startdata
G29.96~-0.02&O6~\tablenotemark{a}&6~\tablenotemark{a}&98~\tablenotemark{b}&0.5~\tablenotemark{c}&-&-&-&-&$2.27''$~\tablenotemark{d}& \\
Model~1&O5.5&6&89&0.11&8.0&1000&20&19.6 & $2.2''$&$50^{\circ}$ \\
Mon~R2~IRS1&O9~\tablenotemark{e}&0.95~\tablenotemark{f}&10.3~\tablenotemark{g}&0.3~\tablenotemark{h}&-&-&-&-&$4.35''$~\tablenotemark{i}& \\
Model~2&B0&0.95&10.3&0.244&3.6&1000&27&6.61 &$4.4''$&$-20^{\circ}$\\
Model~3&B0&0.95&10.3&0.238&8.0&1000&27&9.95 &$6.8''$&$-20^{\circ}$\\
\enddata
\tablecomments{
We estimated the spectral type of the central star for our models 
from our [Ne~II] 
line luminosity, assuming optically thin in the whole region. 
Free parameters of our models are the cloud LSR velocity ($V_{LSR}$),
the ambient medium density ($n_a$), 
the stellar mass loss rate ($\dot{M}_{w}$), the stellar wind speed ($v_{w}$), 
the stellar
speed relative to the ambient medium ($v_{a}$) and the tilt angle.
For models, the size on the sky is the distance from the emission 
peak to the ionizing star after the tilting 
and the beam convolution.
Positive tilt angles represent tilting the symmetry axis away 
from the observers.
The parameters are chosen so that the pressure gradient 
acceleration effect is negligible.} 
\tablenotetext{a}{\cite{praMB99}}
\tablenotetext{b}{\cite{affchk94}}
\tablenotetext{c}{\cite{morSMPDBCR02}}
\tablenotetext{d}{\cite{marBKTH03}}
\tablenotetext{e}{\cite{dowWGJ75}}
\tablenotetext{f}{\cite{racV70}}
\tablenotetext{g}{\cite{meyL91}}
\tablenotetext{h}{\cite{monBMTG90}}
\tablenotetext{i}{\cite{yaohinoswy97}}
\end{deluxetable}









\begin{thebibliography}{57}
\expandafter\ifx\csname natexlab\endcsname\relax\def\natexlab#1{#1}\fi

\bibitem[{{Afflerbach} {et~al.}(1994){Afflerbach}, {Churchwell}, {Hofner}, \&
  {Kurtz}}]{affchk94}
{Afflerbach}, A., {Churchwell}, E., {Hofner}, P., \& {Kurtz}, S. 1994, \apj,
  437, 697

\bibitem[{{Araya} {et~al.}(2002){Araya}, {Hofner}, {Churchwell}, \&
  {Kurtz}}]{araHCK02}
{Araya}, E., {Hofner}, P., {Churchwell}, E., \& {Kurtz}, S. 2002, \apjs, 138,
  63

\bibitem[{{Beck} {et~al.}(1981){Beck}, {Lacy}, {Townes}, {Aller}, {Geballe}, \&
  {Baas}}]{becLTAGB81}
{Beck}, S.~C., {Lacy}, J.~H., {Townes}, C.~H., {Aller}, L.~H., {Geballe},
  T.~R., \& {Baas}, F. 1981, \apj, 249, 592

\bibitem[{{Blaauw}(1993)}]{bla93}
{Blaauw}, A. 1993, in ASP Conf. Ser. 35: Massive Stars: Their Lives in the
  Interstellar Medium, 207

\bibitem[{{Bodenheimer} {et~al.}(1979){Bodenheimer}, {Tenorio-Tagle}, \&
  {Yorke}}]{bodTY79}
{Bodenheimer}, P., {Tenorio-Tagle}, G., \& {Yorke}, H.~W. 1979, \apj, 233, 85

\bibitem[{{Cesaroni} {et~al.}(1992){Cesaroni}, {Walmsley}, \&
  {Churchwell}}]{cesWC92}
{Cesaroni}, R., {Walmsley}, C.~M., \& {Churchwell}, E. 1992, \aap, 256, 618

\bibitem[{{Churchwell}(1990)}]{chu90}
{Churchwell}, E. 1990, \aapr, 2, 79

\bibitem[{{Churchwell} {et~al.}(1990){Churchwell}, {Walmsley}, \&
  {Cesaroni}}]{chuWC90}
{Churchwell}, E., {Walmsley}, C.~M., \& {Cesaroni}, R. 1990, \aaps, 83, 119

\bibitem[{{Comeron}(1997)}]{com97}
{Comeron}, F. 1997, \aap, 326, 1195

\bibitem[{{Comeron} \& {Kaper}(1998)}]{comK98}
{Comeron}, F. \& {Kaper}, L. 1998, \aap, 338, 273

\bibitem[{{De Pree} {et~al.}(2004){De Pree}, {Wilner}, {Mercer}, {Davis},
  {Goss}, \& {Kurtz}}]{depWMDGK04}
{De Pree}, C.~G., {Wilner}, D.~J., {Mercer}, A.~J., {Davis}, L.~E., {Goss},
  W.~M., \& {Kurtz}, S. 2004, \apj, 600, 286

\bibitem[{{Dgani} {et~al.}(1996){Dgani}, {van Buren}, \&
  {Noriega-Crespo}}]{dgaVN96}
{Dgani}, R., {van Buren}, D., \& {Noriega-Crespo}, A. 1996, \apj, 461, 372

\bibitem[{{Downes} {et~al.}(1975){Downes}, {Winnberg}, {Goss}, \&
  {Johansson}}]{dowWGJ75}
{Downes}, D., {Winnberg}, A., {Goss}, W.~M., \& {Johansson}, L.~E.~B. 1975,
  \aap, 44, 243

\bibitem[{{Dyson} {et~al.}(1995){Dyson}, {Williams}, \& {Redman}}]{dysWR95}
{Dyson}, J.~E., {Williams}, R.~J.~R., \& {Redman}, M.~P. 1995, \mnras, 277, 700

\bibitem[{{Fey} {et~al.}(1995){Fey}, {Gaume}, {Claussen}, \& {Vrba}}]{feyGCV95}
{Fey}, A.~L., {Gaume}, R.~A., {Claussen}, M.~J., \& {Vrba}, F.~J. 1995, \apj,
  453, 308

\bibitem[{{Garay} \& {Lizano}(1999)}]{garL99}
{Garay}, G. \& {Lizano}, S. 1999, \pasp, 111, 1049

\bibitem[{{Garay} {et~al.}(1985){Garay}, {Reid}, \& {Moran}}]{garRM85}
{Garay}, G., {Reid}, M.~J., \& {Moran}, J.~M. 1985, \apj, 289, 681

\bibitem[{{Garay} {et~al.}(1993){Garay}, {Rodriguez}, {Moran}, \&
  {Churchwell}}]{garRMC93}
{Garay}, G., {Rodriguez}, L.~F., {Moran}, J.~M., \& {Churchwell}, E. 1993,
  \apj, 418, 368

\bibitem[{{Hanson} {et~al.}(2002){Hanson}, {Luhman}, \& {Rieke}}]{hanLR02}
{Hanson}, M.~M., {Luhman}, K.~L., \& {Rieke}, G.~H. 2002, \apjs, 138, 35

\bibitem[{{Hollenbach} {et~al.}(1994){Hollenbach}, {Johnstone}, {Lizano}, \&
  {Shu}}]{holJLS94}
{Hollenbach}, D., {Johnstone}, D., {Lizano}, S., \& {Shu}, F. 1994, \apj, 428,
  654

\bibitem[{{Hueckstaedt}(2003)}]{hue03}
{Hueckstaedt}, R.~M. 2003, New Astronomy, 8, 295

\bibitem[{{Hughes} \& {Viner}(1976)}]{hugV76}
{Hughes}, V.~A. \& {Viner}, M.~R. 1976, \apj, 204, 55

\bibitem[{{Jaffe} \& {Mart{\'{\i}}n-Pintado}(1999)}]{jafM99}
{Jaffe}, D.~T. \& {Mart{\'{\i}}n-Pintado}, J. 1999, \apj, 520, 162

\bibitem[{{Jaffe} {et~al.}(2003){Jaffe}, {Zhu}, {Lacy}, \&
  {Richter}}]{jafZLR03}
{Jaffe}, D.~T., {Zhu}, Q., {Lacy}, J.~H., \& {Richter}, M. 2003, \apj, 596,
  1053

\bibitem[{{Jones} \& {Walker}(1988)}]{jonW88}
{Jones}, B.~F. \& {Walker}, M.~F. 1988, \aj, 95, 1755

\bibitem[{{Kaper} {et~al.}(1997){Kaper}, {van Loon}, {Augusteijn},
  {Goudfrooij}, {Patat}, {Waters}, \& {Zijlstra}}]{kapVAGPWZ97}
{Kaper}, L., {van Loon}, J.~T., {Augusteijn}, T., {Goudfrooij}, P., {Patat},
  F., {Waters}, L.~B.~F.~M., \& {Zijlstra}, A.~A. 1997, \apjl, 475, L37

\bibitem[{{Kim} \& {Koo}(2001)}]{kimK01}
{Kim}, K. \& {Koo}, B. 2001, \apj, 549, 979

\bibitem[{{Kurtz} {et~al.}(1994){Kurtz}, {Churchwell}, \& {Wood}}]{kurCW94}
{Kurtz}, S., {Churchwell}, E., \& {Wood}, D.~O.~S. 1994, \apjs, 91, 659

\bibitem[{{Lacy} {et~al.}(1982){Lacy}, {Beck}, \& {Geballe}}]{lacBG82}
{Lacy}, J.~H., {Beck}, S.~C., \& {Geballe}, T.~R. 1982, \apj, 255, 510

\bibitem[{{Lacy} {et~al.}(2002){Lacy}, {Richter}, {Greathouse}, {Jaffe}, \&
  {Zhu}}]{lacRGJZ02}
{Lacy}, J.~H., {Richter}, M.~J., {Greathouse}, T.~K., {Jaffe}, D.~T., \& {Zhu},
  Q. 2002, \pasp, 114, 153

\bibitem[{{Lumsden} \& {Hoare}(1999)}]{lumH99}
{Lumsden}, S.~L. \& {Hoare}, M.~G. 1999, \mnras, 305, 701

\bibitem[{{Mac Low} {et~al.}(1991){Mac Low}, {van Buren}, {Wood}, \&
  {Churchwell}}]{macVW91}
{Mac Low}, M., {van Buren}, D., {Wood}, D.~O.~S., \& {Churchwell}, E. 1991,
  \apj, 369, 395

\bibitem[{{Mart{\'{\i}}n-Hern{\' a}ndez} {et~al.}(2003){Mart{\'{\i}}n-Hern{\'
  a}ndez}, {Bik}, {Kaper}, {Tielens}, \& {Hanson}}]{marBKTH03}
{Mart{\'{\i}}n-Hern{\' a}ndez}, N.~L., {Bik}, A., {Kaper}, L., {Tielens},
  A.~G.~G.~M., \& {Hanson}, M.~M. 2003, \aap, 405, 175

\bibitem[{{Meyers-Rice} \& {Lada}(1991)}]{meyL91}
{Meyers-Rice}, B.~A. \& {Lada}, C.~J. 1991, \apj, 368, 445

\bibitem[{{Miralles} {et~al.}(1994){Miralles}, {Rodriguez}, \&
  {Scalise}}]{mirRS94}
{Miralles}, M.~P., {Rodriguez}, L.~F., \& {Scalise}, E. 1994, \apjs, 92, 173

\bibitem[{{Montalban} {et~al.}(1990){Montalban}, {Bachiller}, {Martin-Pintado},
  {Tafalla}, \& {Gomez-Gonzalez}}]{monBMTG90}
{Montalban}, J., {Bachiller}, R., {Martin-Pintado}, J., {Tafalla}, M., \&
  {Gomez-Gonzalez}, J. 1990, \aap, 233, 527

\bibitem[{{Morisset} {et~al.}(2002){Morisset}, {Schaerer},
  {Mart{\'{\i}}n-Hern{\' a}ndez}, {Peeters}, {Damour}, {Baluteau}, {Cox}, \&
  {Roelfsema}}]{morSMPDBCR02}
{Morisset}, C., {Schaerer}, D., {Mart{\'{\i}}n-Hern{\' a}ndez}, N.~L.,
  {Peeters}, E., {Damour}, F., {Baluteau}, J.-P., {Cox}, P., \& {Roelfsema}, P.
  2002, \aap, 386, 558

\bibitem[{{Noriega-Crespo} {et~al.}(1997){Noriega-Crespo}, {van Buren}, \&
  {Dgani}}]{norVD97}
{Noriega-Crespo}, A., {van Buren}, D., \& {Dgani}, R. 1997, \aj, 113, 780

\bibitem[{{Olmi} \& {Cesaroni}(1999)}]{olmC99}
{Olmi}, L. \& {Cesaroni}, R. 1999, \aap, 352, 266

\bibitem[{{Pratap} {et~al.}(1999){Pratap}, {Megeath}, \& {Bergin}}]{praMB99}
{Pratap}, P., {Megeath}, S.~T., \& {Bergin}, E.~A. 1999, \apj, 517, 799

\bibitem[{{Racine} \& {van den Bergh}(1970)}]{racV70}
{Racine}, R. \& {van den Bergh}, S. 1970, in IAU Symp. 38: The Spiral Structure
  of our Galaxy, 219

\bibitem[{{Redman} {et~al.}(1996){Redman}, {Williams}, \& {Dyson}}]{redWD96}
{Redman}, M.~P., {Williams}, R.~J.~R., \& {Dyson}, J.~E. 1996, \mnras, 280, 661

\bibitem[{{Redman} {et~al.}(1998){Redman}, {Williams}, \& {Dyson}}]{redWD98}
---. 1998, \mnras, 298, 33

\bibitem[{{Sewilo} {et~al.}(2004){Sewilo}, {Churchwell}, {Kurtz}, {Goss}, \&
  {Hofner}}]{sewCKGH04}
{Sewilo}, M., {Churchwell}, E., {Kurtz}, S., {Goss}, W.~M., \& {Hofner}, P.
  2004, \apj, 605, 285

\bibitem[{{Takahashi} {et~al.}(2000){Takahashi}, {Matsuhara}, {Watarai}, \&
  {Matsumoto}}]{takMWM00}
{Takahashi}, H., {Matsuhara}, H., {Watarai}, H., \& {Matsumoto}, T. 2000, \apj,
  541, 779

\bibitem[{{Tenorio-Tagle}(1979)}]{ten79}
{Tenorio-Tagle}, G. 1979, \aap, 71, 59

\bibitem[{{Tian} {et~al.}(1996){Tian}, {van Leeuwen}, {Zhao}, \&
  {Su}}]{tiaVZS96}
{Tian}, K.~P., {van Leeuwen}, F., {Zhao}, J.~L., \& {Su}, C.~G. 1996, \aaps,
  118, 503

\bibitem[{{van Buren} \& {Mac Low}(1992)}]{vanM92}
{van Buren}, D. \& {Mac Low}, M. 1992, \apj, 394, 534

\bibitem[{{van Buren} {et~al.}(1990){van Buren}, {Mac Low}, {Wood}, \&
  {Churchwell}}]{vanMWC90}
{van Buren}, D., {Mac Low}, M., {Wood}, D.~O.~S., \& {Churchwell}, E. 1990,
  \apj, 353, 570

\bibitem[{{van Buren} {et~al.}(1995){van Buren}, {Noriega-Crespo}, \&
  {Dgani}}]{vanND95}
{van Buren}, D., {Noriega-Crespo}, A., \& {Dgani}, R. 1995, \aj, 110, 2914

\bibitem[{{Vishniac}(1994)}]{vis94}
{Vishniac}, E.~T. 1994, \apj, 428, 186

\bibitem[{{Wilkin}(1996)}]{wil96}
{Wilkin}, F.~P. 1996, \apjl, 459, L31

\bibitem[{{Williams} {et~al.}(1996){Williams}, {Dyson}, \& {Redman}}]{wilDR96}
{Williams}, R.~J.~R., {Dyson}, J.~E., \& {Redman}, M.~P. 1996, \mnras, 280, 667

\bibitem[{{Wood} \& {Churchwell}(1989{\natexlab{a}})}]{wooC89a}
{Wood}, D.~O.~S. \& {Churchwell}, E. 1989{\natexlab{a}}, \apj, 340, 265

\bibitem[{{Wood} \& {Churchwell}(1989{\natexlab{b}})}]{wooC89b}
---. 1989{\natexlab{b}}, \apjs, 69, 831

\bibitem[{{Wood} \& {Churchwell}(1991)}]{wooC91}
---. 1991, \apj, 372, 199

\bibitem[{{Yao} {et~al.}(1997){Yao}, {Hirata}, {Ishii}, {Nagata}, {Ogawa},
  {Sato}, {Watanabe}, \& {Yamashita}}]{yaohinoswy97}
{Yao}, Y., {Hirata}, N., {Ishii}, M., {Nagata}, T., {Ogawa}, Y., {Sato}, S.,
  {Watanabe}, M., \& {Yamashita}, T. 1997, \apj, 490, 281

\bibitem[{{Yorke} {et~al.}(1983){Yorke}, {Tenorio-Tagle}, \&
  {Bodenheimer}}]{yorTB83}
{Yorke}, H.~W., {Tenorio-Tagle}, G., \& {Bodenheimer}, P. 1983, \aap, 127, 313

\end{thebibliography}
\end{document}